\documentclass[aps,prd,twocolumn,superscriptaddress,amsmath,amssymb,showpacs]{revtex4}
\usepackage{xspace}
\usepackage{subfigure}
\usepackage{times}
\usepackage{graphicx} 
\usepackage{dcolumn} 
\usepackage{epstopdf}
\usepackage{bm}        % for math
\usepackage{amssymb}   % for math
                        
\DeclareGraphicsExtensions{.pdf,.eps,.png,.jpg,.jpeg}
\graphicspath{{figures/}}

\newcommand{\ppbar}{\ensuremath{p\overline{p}}\xspace}

\def\GeVc{GeV$\!/c$}
\def\GeVcc{GeV$\!/c^2$}

\def\mMEt{\not\kern-.35em {E_T}}

\newcommand{\mett}{\mbox{$E\!\!\!\!/_{T}$}\xspace}

\def\simle{\mathrel{
   \rlap{\raise 0.511ex \hbox{$<$}}{\lower 0.511ex \hbox{$\sim$}}}}

\begin{document}
%%%%%%%%%%%%%%%%%%%%%%%%%%%%%%%%%%%%%%%%%%%%%%%
% Toggle double spacing and line numbering
% Won't work with the PRD revtex4 !
%\pagewiselinenumbers % collab
%\linenumbers
%\doublespace
%%%%%%%%%%%%%%%%%%%%%%%%%%%%%%%%%%%%%%%%%%%%%%%
% the following line is for submission, including submission to the arXiv!!

%\mbox{FERMILAB-PUB-XX-YYY-Z}
%\hfill\mbox{CDF/PHYS/ELECTROWEAK/PUBLIC/10311}
%\preprint{FERMILAB-PUB-11-375-E-PPD}
%\preprint{CDF/PHYS/ELECTROWEAK/PUBLIC/xxxx}
%\hfill\mbox{Version 4.0}

%\input{phys_defs.tex} 

\title{Search for the production of \textit{ZW} and \textit{ZZ} boson pairs decaying into charged leptons and jets in \ppbar collisions at $\sqrt{s}=1.96$~TeV}

%\author{The CDF Collaboration}
%\affiliation{\url{http://www-cdf.fnal.gov}}
% Last update: $Date: 2013/09/25 17:11:00 $
\affiliation{Institute of Physics, Academia Sinica, Taipei, Taiwan 11529, Republic of China}
\affiliation{Argonne National Laboratory, Argonne, Illinois 60439, USA}
\affiliation{University of Athens, 157 71 Athens, Greece}
\affiliation{Institut de Fisica d'Altes Energies, ICREA, Universitat Autonoma de Barcelona, E-08193, Bellaterra (Barcelona), Spain}
\affiliation{Baylor University, Waco, Texas 76798, USA}
\affiliation{Istituto Nazionale di Fisica Nucleare Bologna, \ensuremath{^{ii}}University of Bologna, I-40127 Bologna, Italy}
\affiliation{University of California, Davis, Davis, California 95616, USA}
\affiliation{University of California, Los Angeles, Los Angeles, California 90024, USA}
\affiliation{Instituto de Fisica de Cantabria, CSIC-University of Cantabria, 39005 Santander, Spain}
\affiliation{Carnegie Mellon University, Pittsburgh, Pennsylvania 15213, USA}
\affiliation{Enrico Fermi Institute, University of Chicago, Chicago, Illinois 60637, USA}
\affiliation{Comenius University, 842 48 Bratislava, Slovakia; Institute of Experimental Physics, 040 01 Kosice, Slovakia}
\affiliation{Joint Institute for Nuclear Research, RU-141980 Dubna, Russia}
\affiliation{Duke University, Durham, North Carolina 27708, USA}
\affiliation{Fermi National Accelerator Laboratory, Batavia, Illinois 60510, USA}
\affiliation{University of Florida, Gainesville, Florida 32611, USA}
\affiliation{Laboratori Nazionali di Frascati, Istituto Nazionale di Fisica Nucleare, I-00044 Frascati, Italy}
\affiliation{University of Geneva, CH-1211 Geneva 4, Switzerland}
\affiliation{Glasgow University, Glasgow G12 8QQ, United Kingdom}
\affiliation{Harvard University, Cambridge, Massachusetts 02138, USA}
\affiliation{Division of High Energy Physics, Department of Physics, University of Helsinki, FIN-00014, Helsinki, Finland; Helsinki Institute of Physics, FIN-00014, Helsinki, Finland}
\affiliation{University of Illinois, Urbana, Illinois 61801, USA}
\affiliation{The Johns Hopkins University, Baltimore, Maryland 21218, USA}
\affiliation{Institut f\"{u}r Experimentelle Kernphysik, Karlsruhe Institute of Technology, D-76131 Karlsruhe, Germany}
\affiliation{Center for High Energy Physics: Kyungpook National University, Daegu 702-701, Korea; Seoul National University, Seoul 151-742, Korea; Sungkyunkwan University, Suwon 440-746, Korea; Korea Institute of Science and Technology Information, Daejeon 305-806, Korea; Chonnam National University, Gwangju 500-757, Korea; Chonbuk National University, Jeonju 561-756, Korea; Ewha Womans University, Seoul, 120-750, Korea}
\affiliation{Ernest Orlando Lawrence Berkeley National Laboratory, Berkeley, California 94720, USA}
\affiliation{University of Liverpool, Liverpool L69 7ZE, United Kingdom}
\affiliation{University College London, London WC1E 6BT, United Kingdom}
\affiliation{Centro de Investigaciones Energeticas Medioambientales y Tecnologicas, E-28040 Madrid, Spain}
\affiliation{Massachusetts Institute of Technology, Cambridge, Massachusetts 02139, USA}
\affiliation{University of Michigan, Ann Arbor, Michigan 48109, USA}
\affiliation{Michigan State University, East Lansing, Michigan 48824, USA}
\affiliation{Institution for Theoretical and Experimental Physics, ITEP, Moscow 117259, Russia}
\affiliation{University of New Mexico, Albuquerque, New Mexico 87131, USA}
\affiliation{The Ohio State University, Columbus, Ohio 43210, USA}
\affiliation{Okayama University, Okayama 700-8530, Japan}
\affiliation{Osaka City University, Osaka 558-8585, Japan}
\affiliation{University of Oxford, Oxford OX1 3RH, United Kingdom}
\affiliation{Istituto Nazionale di Fisica Nucleare, Sezione di Padova, \ensuremath{^{jj}}University of Padova, I-35131 Padova, Italy}
\affiliation{University of Pennsylvania, Philadelphia, Pennsylvania 19104, USA}
\affiliation{Istituto Nazionale di Fisica Nucleare Pisa, \ensuremath{^{kk}}University of Pisa, \ensuremath{^{ll}}University of Siena, \ensuremath{^{mm}}Scuola Normale Superiore, I-56127 Pisa, Italy, \ensuremath{^{nn}}INFN Pavia, I-27100 Pavia, Italy, \ensuremath{^{oo}}University of Pavia, I-27100 Pavia, Italy}
\affiliation{University of Pittsburgh, Pittsburgh, Pennsylvania 15260, USA}
\affiliation{Purdue University, West Lafayette, Indiana 47907, USA}
\affiliation{University of Rochester, Rochester, New York 14627, USA}
\affiliation{The Rockefeller University, New York, New York 10065, USA}
\affiliation{Istituto Nazionale di Fisica Nucleare, Sezione di Roma 1, \ensuremath{^{pp}}Sapienza Universit\`{a} di Roma, I-00185 Roma, Italy}
\affiliation{Mitchell Institute for Fundamental Physics and Astronomy, Texas A\&M University, College Station, Texas 77843, USA}
\affiliation{Istituto Nazionale di Fisica Nucleare Trieste, \ensuremath{^{qq}}Gruppo Collegato di Udine, \ensuremath{^{rr}}University of Udine, I-33100 Udine, Italy, \ensuremath{^{ss}}University of Trieste, I-34127 Trieste, Italy}
\affiliation{University of Tsukuba, Tsukuba, Ibaraki 305, Japan}
\affiliation{Tufts University, Medford, Massachusetts 02155, USA}
\affiliation{University of Virginia, Charlottesville, Virginia 22906, USA}
\affiliation{Waseda University, Tokyo 169, Japan}
\affiliation{Wayne State University, Detroit, Michigan 48201, USA}
\affiliation{University of Wisconsin, Madison, Wisconsin 53706, USA}
\affiliation{Yale University, New Haven, Connecticut 06520, USA}

\author{T.~Aaltonen}
\affiliation{Division of High Energy Physics, Department of Physics, University of Helsinki, FIN-00014, Helsinki, Finland; Helsinki Institute of Physics, FIN-00014, Helsinki, Finland}
\author{S.~Amerio\ensuremath{^{jj}}}
\affiliation{Istituto Nazionale di Fisica Nucleare, Sezione di Padova, \ensuremath{^{jj}}University of Padova, I-35131 Padova, Italy}
\author{D.~Amidei}
\affiliation{University of Michigan, Ann Arbor, Michigan 48109, USA}
\author{A.~Anastassov\ensuremath{^{v}}}
\affiliation{Fermi National Accelerator Laboratory, Batavia, Illinois 60510, USA}
\author{A.~Annovi}
\affiliation{Laboratori Nazionali di Frascati, Istituto Nazionale di Fisica Nucleare, I-00044 Frascati, Italy}
\author{J.~Antos}
\affiliation{Comenius University, 842 48 Bratislava, Slovakia; Institute of Experimental Physics, 040 01 Kosice, Slovakia}
\author{G.~Apollinari}
\affiliation{Fermi National Accelerator Laboratory, Batavia, Illinois 60510, USA}
\author{J.A.~Appel}
\affiliation{Fermi National Accelerator Laboratory, Batavia, Illinois 60510, USA}
\author{T.~Arisawa}
\affiliation{Waseda University, Tokyo 169, Japan}
\author{A.~Artikov}
\affiliation{Joint Institute for Nuclear Research, RU-141980 Dubna, Russia}
% ...

\collaboration{CDF Collaboration}
\altaffiliation[With visitors from]{
\ensuremath{^{a}}University of British Columbia, Vancouver, BC V6T 1Z1, Canada,
\ensuremath{^{b}}Istituto Nazionale di Fisica Nucleare, Sezione di Cagliari, 09042 Monserrato (Cagliari), Italy,
\ensuremath{^{c}}University of California Irvine, Irvine, CA 92697, USA,
\ensuremath{^{d}}Institute of Physics, Academy of Sciences of the Czech Republic, 182~21, Czech Republic,
\ensuremath{^{e}}CERN, CH-1211 Geneva, Switzerland,
\ensuremath{^{f}}Cornell University, Ithaca, NY 14853, USA,
\ensuremath{^{g}}University of Cyprus, Nicosia CY-1678, Cyprus,
\ensuremath{^{h}}Office of Science, U.S. Department of Energy, Washington, DC 20585, USA,
\ensuremath{^{i}}University College Dublin, Dublin 4, Ireland,
\ensuremath{^{j}}ETH, 8092 Z\"{u}rich, Switzerland,
\ensuremath{^{k}}University of Fukui, Fukui City, Fukui Prefecture, Japan 910-0017,
\ensuremath{^{l}}Universidad Iberoamericana, Lomas de Santa Fe, M\'{e}xico, C.P. 01219, Distrito Federal,
\ensuremath{^{m}}University of Iowa, Iowa City, IA 52242, USA,
\ensuremath{^{n}}Kinki University, Higashi-Osaka City, Japan 577-8502,
\ensuremath{^{o}}Kansas State University, Manhattan, KS 66506, USA,
\ensuremath{^{p}}Brookhaven National Laboratory, Upton, NY 11973, USA,
\ensuremath{^{q}}Queen Mary, University of London, London, E1 4NS, United Kingdom,
\ensuremath{^{r}}University of Melbourne, Victoria 3010, Australia,
\ensuremath{^{s}}Muons, Inc., Batavia, IL 60510, USA,
\ensuremath{^{t}}Nagasaki Institute of Applied Science, Nagasaki 851-0193, Japan,
\ensuremath{^{u}}National Research Nuclear University, Moscow 115409, Russia,
\ensuremath{^{v}}Northwestern University, Evanston, IL 60208, USA,
\ensuremath{^{w}}University of Notre Dame, Notre Dame, IN 46556, USA,
\ensuremath{^{x}}Universidad de Oviedo, E-33007 Oviedo, Spain,
\ensuremath{^{y}}CNRS-IN2P3, Paris, F-75205 France,
\ensuremath{^{z}}Universidad Tecnica Federico Santa Maria, 110v Valparaiso, Chile,
\ensuremath{^{aa}}The University of Jordan, Amman 11942, Jordan,
\ensuremath{^{bb}}Universite catholique de Louvain, 1348 Louvain-La-Neuve, Belgium,
\ensuremath{^{cc}}University of Z\"{u}rich, 8006 Z\"{u}rich, Switzerland,
\ensuremath{^{dd}}Massachusetts General Hospital, Boston, MA 02114 USA,
\ensuremath{^{ee}}Harvard Medical School, Boston, MA 02114 USA,
\ensuremath{^{ff}}Hampton University, Hampton, VA 23668, USA,
\ensuremath{^{gg}}Los Alamos National Laboratory, Los Alamos, NM 87544, USA,
\ensuremath{^{hh}}Universit\`{a} degli Studi di Napoli Federico I, I-80138 Napoli, Italy
}
\noaffiliation
% Last update: $Date: 2013/09/25 17:11:00 $

%\date{June 24, 2012}
%\date{September 7, 2012}
%\date{March 7, 2013}
%\date{July 9, 2013}
%\date{September 10, 2013}
\date{September 30, 2013}

\begin{abstract}
  % remove the space for publication
  %\vspace*{3.0cm} 
%Diboson production where $WW$ is a significant component has been observed at the Tevatron collider in semi-hadronic decay modes.
We present a measurement of the production cross section for \textit{ZW} and \textit{ZZ} boson pairs in final states with a pair of charged leptons, from the decay of a \textit{Z} boson, and at least two jets, from the decay of a \textit{W} or \textit{Z} boson, using the full sample of proton-antiproton collisions recorded with the CDF~II detector at the Tevatron, corresponding to 8.9 fb$^{-1}$ of integrated luminosity.  We increase the sensitivity to vector boson decays into pairs of quarks using a neural network discriminant that exploits the differences between the spatial spread of energy depositions and charged-particle momenta contained within the jet of particles originating from quarks and gluons. Additionally, we employ new jet energy corrections to Monte Carlo simulations that account for differences in the observed energy scales for quark and gluon jets. The number of signal events is extracted through a simultaneous fit to the dijet mass spectrum in three classes of events: events likely to contain jets with a heavy-quark decay, events likely to contain jets originating from light quarks, and events that fail these identification criteria. We determine the production cross section to be $\sigma_{ZW+ZZ} = 2.5^{+2.0}_{-1.0}$~pb ($< 6.1$~pb at the 95\% confidence level), consistent with the standard model prediction of $5.1$~pb.

\end{abstract}
  
% activate the following line for publication
%\pacs{ 14.80.Bn, 14.70.-e, 12.15.-y}

\maketitle

\section{Introduction}
\label{sec:Intro}
The standard model (SM) offers precise predictions of production rates associated with self-interactions of the gauge bosons~\cite{campbell}. Differences between these predictions and measured diboson production cross sections may indicate the presence of non-SM physics~\cite{hagiwara,kober}, even specifically in hadronic final states~\cite{technicolor}. Additionally, since hadronic final states in diboson production are similar to those from associated Higgs boson production ($p\bar{p}\to VH+X$\ where $V$=$W,Z$), the analysis techniques used to measure diboson production in partially hadronic final states are relevant to searches for associated Higgs boson production.

Measurements of diboson production are typically difficult due to the small production cross sections of the order of 10 pb or less~\cite{campbell}.  Furthermore, measurements of decay channels where one $W$ or $Z$ boson decays hadronically are particularly challenging at hadron colliders: although expected event yields are larger than those in purely leptonic decay channels due to the higher hadronic decay ($V \rightarrow q\bar{q'}$) branching ratio, the expected backgrounds from QCD multijet processes and $V$+ jets production are also much greater. Experiments at the Tevatron have previously measured the cross sections for pair production of gauge bosons in partially hadronic decay channels~\cite{metjj_prl,lmetjj_prl_ME,lmetjj_prl_Mjj,D0_lmetjj}, but all of these measurements have included contributions from \textit{WW} production, which has a higher cross section than that for combined \textit{ZW} and \textit{ZZ} production. Searches using identification of $b$-quark decays in the final states ($b$-tagging) to increase sensitivity to events with $Z \rightarrow b\bar{b}$ decays have been performed~\cite{metbb_prd}, but have not yet provided observations of $ZV$ production in partially hadronic decay channels.

We present a study of $ZV$ production from a final state with two leptons and at least two jets~\cite{Ketchum_thesis_prl}. We require the two leptons to originate from the decay of a $Z$ boson and search for associated $V \rightarrow q\bar{q'}$ decays by performing a fit to the dijet invariant mass ($m_{jj}$) spectrum of the two leading-$E_{T}$~\cite{coordinateSystem} jets. To maximize sensitivity to diboson production, we separate events into three channels: a heavy-flavor-tagged channel, largely sensitive to $ZZ \rightarrow \ell^{+}\ell^{-}b\bar{b}$ decays; a light-flavor-tagged channel, which uses a new artificial-neural-network-based discriminant to preferentially select events with quark-like jets over gluon-like jets; and an untagged channel, which contains the remaining events that pass the event selection requirements. The final fit to the $m_{jj}$ spectra is performed simultaneously across all three channels.

This paper is organized as follows: in Sec.~\ref{sec:Detector} we briefly describe the CDF~II detector; in Sec.~\ref{sec:selection} we describe the data sets and event selection requirements that are used in the $ZV$ search; in Sec.~\ref{sec:jesQG} we show the derivation of new jet energy corrections to Monte Carlo simulations that account for differences in the observed energy scales of quark and gluon jets; in Sec.~\ref{sec:QGvalue} we provide details of a new neural network-based discriminant that identifies jets more likely to originate from quarks than from gluons; and, in Sec.~\ref{sec:extraction}, we describe the signal-extraction method, and report the results of the $ZV$ search.

\section{The CDF~II detector}
\label{sec:Detector}
The CDF~II detector is described in detail elsewhere~\cite{CDF_detect_A}. The detector is cylindrically symmetric around the Tevatron beam line. Tracking detectors are installed around the interaction point to reconstruct the trajectories of charged particles (tracks). The tracking systems are located within a superconducting solenoid that produces a $1.4$~T magnetic field aligned with the $\ppbar$ beams. Around the outside of the solenoid, calorimeter modules arranged in a projective-tower geometry measure the energies of charged and neutral particles. Drift chambers outside the calorimeter are used to detect muons, which typically deposit little energy in the calorimeter.

%\subsection{Tracking}
%\label{sec:tracking}
The central outer tracker (COT) is a $3.1$ m long open-cell drift chamber that has 96 measurement layers in the region between $0.40$ and $1.37$ m from the beam axis, providing full track coverage in the pseudorapidity region $|\eta| < 1.0$. Sense wires are arranged in eight alternating axial and $\pm~2^{\circ}$ stereo ``superlayers" with 12 wires each. The position resolution of a single drift-time measurement is about $140~\mu$m.
%Charged particle trajectories are found initially as a series of approximate line segments in individual axial superlayers. Two complementary algorithms associate segments lying on a common circle, and the results are merged to form a final set of axial tracks segments.
%Tracks are reconstructed in three dimensions by associating track segments in stereo superlayers with the axial track segments.
%The efficiency of finding isolated high-momentum tracks is measured using electrons from $W^{\pm} \rightarrow e^{\pm}\nu$ decays identified in the central region $|\eta| \le 1.1$ using only calorimetric information from the electron shower and the missing transverse energy. The efficiency for finding these electron tracks is $99.93^{+0.07}_{-0.35}\%$, and this is typical for isolated high-momentum tracks from from either electronic or muonic $W$ and $Z$ decays contained in the COT.  The transverse momentum resolution of high-$p_{T}$ tracks is $\delta p_{T}/p_{T}^{2} \approx 0.1\%~($\GeVc$)^{-1}$.
%Their track position resolution in the direction along the beam line at the origin is $\delta z \approx 0.5$\,cm, and the resolution on the track impact parameter, the distance from the beam line to the track's closest approach in the transverse plane, is $\delta d_{0} \approx 350~\mu$m.
A five-layer double-sided silicon microstrip detector (SVX) covers the region between $2.5$ to $11$ cm from the beam axis. Three separate SVX barrel modules along the beam line cover a length of 96 cm, approximately 90\% of the luminous beam interaction region. Three of the five layers combine an $r$-$\phi$ measurement on one side and a $90^{\circ}$ stereo measurement on the other, and the remaining two layers combine an $r$-$\phi$ measurement with a small angle ($\pm 1.2^{\circ}$) stereo measurement. The typical silicon hit resolution is 11~$\mu$m. An intermediate silicon layers at a radius of 22 cm from the beam line in the central region links tracks in the COT to hits in the SVX. The fiducial range of the silicon detector extends to pseudorapidity magnitude $|\eta| < 2.0$.
%Silicon hit information is added to COT tracks using a progressing ``outside-in" tracking algorithm in which COT tracks are extrapolated into the silicon detector, associated silicon hits are found, and the track is refit with the added information of the silicon measurements.
%The initial track parameters provide a width for a search road in a given layer. Then, for each candidate hit in that layer, the track is refit and used to define the search road into the next layer. This stepwise addition of precision SVX information at each layer progressively reduces the size of the search road, while also accounting for the additional uncertainty due to multiple scattering in each layer. The search uses the two best candidate hits in each layer to generate a small tree of final track candidates, from which the tracks with the best $\chi^{2}$ are selected.
%The efficiency for associating at least three silicon hits with an isolated COT track is $91 \pm 1\%$.
%The extrapolated impact parameter resolution for high-momentum outside-in tracks is much smaller than for COT-only tracks: $30~\mu$m, including the uncertainty in the beam position.

%\subsection{Calorimetry}
%\label{sec:calorimetry}

Calorimeter modules are located outside the central tracking volume and solenoid. The inner electromagnetic layers consist of lead sheets interspersed with scintillators, while the outer hadronic layers consist of scintillators sandwiched between steel sheets. The calorimeter is split between central barrel ($|\eta| < 1.1$) and forward end-plug ($1.1 < |\eta| < 3.6$) sections. Individual towers in the central barrel subtend $0.1$ in $|\eta|$ and $15^{\circ}$ in $\phi$. The sizes of the towers in the end plug calorimeter vary with $|\eta|$, subtending $0.1$ in $|\eta|$ and $7.5^{\circ}$ in $\phi$ at $|\eta|=1.1$, and $0.5$ in $|\eta|$ and $15^{\circ}$ in $\phi$ at $|\eta|=3.6$. The energy resolution in the electromagnetic calorimeters is $14\%/\sqrt{E_{T}}$ in the central barrel and $16\%/\sqrt{E} \oplus 1\%$ in the forward end-plug section, with the energies in units of GeV. The single-particle energy resolution in the hadronic calorimeters, measured using pions, ranges from $75\%/\sqrt{E}$ in the central barrel to $80\%/\sqrt{E} \oplus 5\%$ in the forward end-plug section, with the energies expressed in units of GeV.

The hadronization of quarks and gluons produced in the interaction leads to collimated groups of high-momentum particles called jets. These jets, along with photons and electrons, leave isolated energy deposits in contiguous groups of calorimeter towers, which can be summed together into an energy cluster. Electrons and photons are identified as isolated, mostly electromagnetic clusters, and quality requirements may be placed on the presence of a high-$p_{T}$ track geometrically matched to the cluster to more accurately identify electrons. Jets are identified as electromagnetic and hadronic clusters with the combined electromagnetic fraction $E_{EM} / E_{total} = E_{EM} / (E_{EM}+E_{had}) < 0.9$, clustered using the {\sc jetclu} cone algorithm~\cite{jetclu} with a fixed cone size of $\Delta R\equiv\sqrt{(\Delta\eta)^{2}+(\Delta\phi)^{2}}=0.4$. 

%\subsection{Muon detectors}
%\label{sec:muon_detectors}
Outside the calorimeters, drift chambers detect muons. A four-layer stack of planar drift chambers detect muons with $p_{T} > 1.4$~\GeVc, and another four layers of drift chambers behind 60 cm of steel detect muons with $p_{T} > 2.0$~\GeVc. Both systems cover a region of $|\eta| < 0.6$, though they have different structure and their geometrical coverages do not overlap exactly. Muons in the region between $0.6 < |\eta| < 1.0$ pass through at least four drift layers arranged within a conic section outside of the central calorimeter. Muons are identified as either COT tracks that extrapolate to hits in the muon detectors, or as isolated tracks unmatched to hits in the muon detectors that satisfy tighter tracking-quality requirements and extrapolate to calorimeter energy depositions consistent with a minimum ionizing particle.

\section{Data set and event selection}
\label{sec:selection}
We analyze the full data set of \ppbar collisions collected by the CDF~II detector. We require events to be collected from periods when the tracking systems, calorimeters, muon detectors were all functioning properly, corresponding to an integrated luminosity of $8.9$~fb$^{-1}$. Events are selected via a number of high-$E_{T}$ electron and high-$p_{T}$ muon online event-selection requirements (triggers). The majority of these triggers require at least one electron (muon) with $E_{T} > 18$~GeV ($p_{T} > 18$~\GeVc). We require the events to contain two electrons (muons) with $E_{T} > 20$~GeV ($p_{T} >20$~\GeVc)~and determine the trigger selection and event reconstruction efficiencies by comparing the number of data and simulated $Z \rightarrow \ell\ell$ events containing exactly one jet with $E_{T} > 20$~GeV.

For the final analysis, we select events with at least two leptons, and two or more jets.  In the unlikely case that more than two charged leptons are reconstructed, we select the two leptons with highest $p_{T}$. In addition to the $p_{T}$ requirements on the leptons, we require leptons associated with well-reconstructed tracks (central electrons, $|\eta|<1$, and all muons) to be of opposite charge, and a reconstructed dilepton invariant mass, $m_{\ell\ell}$, consistent with the mass of the $Z$ boson, $76 < m_{\ell\ell} < 106$~\GeVcc. We require both leading-$E_{T}$ jets to have $E_{T} > 25$~GeV and $|\eta| < 2.0$, and to be spatially separated from the reconstructed leptons ($\Delta R >0.4$). Additionally, the two jets must be separated by $\Delta R >0.7$.  Finally, as the final state should contain no particles that are not reconstructed in the detector, we also require that the missing transverse energy, $\mett$~\cite{metdef}, is less than  $20$~GeV. 

After this selection, three major sources of background contribute. The dominant background comes from the production of a $Z$ boson, decaying to an $e^{+}e^{-}$ or $\mu^{+}\mu^{-}$ pair, in association with two jets. Simulated events generated using {\sc alpgen}~\cite{alpgen}, interfaced with {\sc pythia}~\cite{pythia} for showering, are used to estimate this background. The production cross sections for $Z+b\bar{b}$ processes are normalized to experimental measurements~\cite{ZbPRD}.

Another significant background results from jets misidentified as leptons. The contributions from these lepton \textit{fakes} are estimated using data-driven methods. For muons, we use events with same-sign muon pairs (rather than opposite-sign) that otherwise satisfy all event selection requirements. For electrons, we derive a misidentification rate representing the likelihood for a jet to be misidentified as an electron, as a function of jet $E_{T}$ and $\eta$, using jet-triggered data with minimal contributions from events with electrons. This rate is then applied to all possible electron-jet paris in events from from the high-$p_{T}$ electron data set, where the jet is then treated as a second electron, and the event selection requirements are otherwise applied normally.

While the requirement to have low \mett reduces its total contribution, top-quark-pair production, where each top quark decays into a leptonic final state ($t\bar{t} \rightarrow W^{+}bW^{-}\bar{b} \rightarrow \ell^{+}\nu_{\ell}b\ell^{-}\bar{\nu_{\ell}}\bar{b}$), contributes events to the final event sample, especially in the heavy-flavor-tagged channel.  We estimate $t\bar{t}$ contributions using {\sc pythia} with $\sigma_{t\bar{t}}=7.5$~pb and $m_{t}=172.5$~\GeVcc. Finally, \textit{ZW} and \textit{ZZ} signal events are also modeled using {\sc pythia}. The predicted and observed numbers of events are shown in Table~\ref{tab:events_predicted}.

\begin{table*}[!hbt]
\begin{center}
\begin{ruledtabular}
\begin{tabular}{lcccc}
& All events & Heavy-flavor-tagged & Light-flavor-tagged & Untagged \\ 
\hline
%$Z$+jets & $8\,667 \pm 1\,113$& $93 \pm 14$& $1\,454 \pm 307$& $6\,721 \pm 968$ \\
%$Z$+$b$ jets & $714 \pm 299$ & $111 \pm  48$ & $53.8 \pm 25.5$ & $536 \pm 230$ \\
%$t\bar{t}$ & $9.2 \pm 0.9$ & $3.3 \pm 0.4$ & $0.7 \pm 0.1$ & $5.2 \pm 0.6$\\
%Misidentified leptons & $330 \pm 165$ & $4.8 \pm 2.4$ & $39.4 \pm 20.3$ & $283 \pm 142$ \\
%\hline
%Predicted background & $9\,720 \pm 1\,247$ & $212 \pm 55$ & $1,617 \pm 325$ & $7\,890 \pm 1\,071$ \\
%$ZW+ZZ$ & $313 \pm 29$ & $12.8 \pm 1.6$ & $84.8 \pm 12.3$& $205 \pm 22$ \\
%\hline
%Total predicted events & $10\,033 \pm 1\,259$ & $225 \pm 55$ & $1\,706 \pm 331$ & $8\,102 \pm 1\,080$ \\
%Data events & $9\,846$ & $172$ & $1\,724$ & $7\,950$ \\ 
$Z$+jets & $8\,700 \pm 1\,100$& $93 \pm 14$& $1\,520 \pm 310$& $7\,100 \pm 970$ \\
$Z$+$b$ jets & $710 \pm 300$ & $111 \pm  48$ & $55 \pm 26$ & $550 \pm 230$ \\
$t\bar{t}$ & $9.2 \pm 0.9$ & $3.3 \pm 0.4$ & $0.7 \pm 0.1$ & $5.1 \pm 0.6$\\
Misidentified leptons & $330 \pm 170$ & $4.8 \pm 2.4$ & $41 \pm 20$ & $280 \pm 140$ \\
\hline
Predicted background & $9\,700 \pm 1\,200$ & $212 \pm 55$ & $1\,620 \pm 330$ & $7\,900 \pm 1\,100$ \\
$ZW+ZZ$ & $313 \pm 29$ & $12.8 \pm 1.6$ & $89 \pm 12$& $212 \pm 22$ \\
\hline
Total predicted events & $10\,000 \pm 1\,300$ & $225 \pm 55$ & $1\,710 \pm 330$ & $8\,100 \pm 1\,100$ \\
Data events & $9\,846$ & $172$ & $1\,724$ & $7\,950$ \\ 
\end{tabular}
\end{ruledtabular}
\caption{Predicted and observed numbers of events in the final event selection, where the numbers of events are rounded to the appropriate significant figures given the uncertainties. The uncertainties incorporate all systematic uncertainties summarized in Table~\ref{tab:systematics} and include an additional 10\% uncertainty on the normalization of $Z+$jets events and a 6\% uncertainty on the normalization of $ZW+ZZ$ events, from the theoretical uncertainties on the production cross sections for those processes.}
\label{tab:events_predicted}
\end{center}
\end{table*}

\section{Jet Energy Calibration}
\label{sec:jesQG}
The energies of jets, measured in the calorimeter, are corrected to account for a number of effects that distort the true jet energy. These effects include changes in calorimeter performance as a function of $|\eta|$ and time, contributions from multiple \ppbar interactions per beam crossing (pileup), contributions from the other partons in the interacting proton and antiproton (underlying event), the non-linear response of the calorimeter, and energy radiated outside of the jet cone. These jet energy scale (JES) corrections are described in detail in Ref.~\cite{cdf_JES}.

These energy corrections, however, do not attempt to account for potential differences in the modeled calorimeter response to jets originating from quarks and gluons. For example, the largest corrections modify the energy scale of the jets to more accurately match that of the initial parton energies and their resulting particle jets, and are derived using {\sc pythia}~\cite{pythia} dijet Monte Carlo (MC) simulations, but independently of the initiating type of parton. Differences in the response of gluon and quark jets between simulation and data lead to differences in the measured energies for these jets that are not covered by the previously-assigned systematic uncertainties on the JES~\cite{cdf_JES_precision_limits}.

We derive a data-driven correction for the response to quark and gluon jets in simulated events using two independent samples of jets with different compositions of quark- and gluon-initiated jets. In these samples, we derive a correction to the jet energy by balancing the transverse energy of the jets against particles of known transverse momentum. We use events where a jet recoils against a high-$E_{T}$ photon---a sample rich in quark jets (based on simulations modeled using {\sc pythia})---and utilize the significant number of $Z \rightarrow \ell^{+}\ell^{-}+$ jet events available in the full CDF data set, which have a larger fraction of gluon jets. The quark and gluon content of these two samples differ due to the difference in mass between the $Z$ boson and the photon: because the $Z$ boson mass is higher, the initial partons of the production process typically carry a higher fraction of the momentum of the proton than those involved in the production of high-energy photons. This leads to a difference in the quark and gluon content of these samples.

To derive a correction, we construct the \textit{balance} of the jet with these more accurately measured reference particles:

\begin{equation}
K_{Z, \gamma} = ( E_{T}\textsuperscript{jet} / p_{T}^{Z, \gamma}) - 1~\text{.}
\label{eq:balance_def}
\end{equation}

For unbiased measurements of the jet energy, the balance will equal zero.  Samples of jets with non-zero balance could be corrected with a jet-energy correction factor of $1/(K_{Z, \gamma} +1)$. However, rather than derive independent JES corrections for quark and gluon jets in data and simulation, we compare the balance in data and simulation and derive an additional correction to be applied to simulated jets, dependent upon whether these jets are matched to quarks or gluons. The correction to simulated quark-jet energies is $+1.4 \pm 2.7\%$, while the correction to gluon-jet energies is much larger: $-7.9 \pm 4.4\%$.

\subsection{Data set and event selection}
\label{sec:jesQG_selection}
The data set and event selection for the $Z$-jet balancing sample largely follow those described in Sec.~\ref{sec:selection}. We require two leptons consistent with resulting from the decay of a $Z$ boson and exactly one jet with $E_{T} > 15$~GeV and no other jets with (uncorrected) $E_{T} > 3$~GeV within $|\eta| < 2.4$. Additionally, we ensure that the $Z$ boson and jet are azimuthally opposite (back-to-back) by requiring their azimuthal separation to exceed $2.8$~radians, and we require that $p_{T}^{Z} > 10$~\GeVc.

For the $\gamma$-jet balancing sample, we closely mirror the selection requirements described in Ref.~\cite{cdf_JES}. We use events collected with an isolated-central-photon trigger over the same period of time as that of the high-$p_{T}$ lepton samples. We compare these data to {\sc pythia} simulations of both $\gamma +$ jet production as well as dijet production, which also contributes to the $\gamma$-jet balancing sample.

To match the selection requirements of the isolated-central-photon trigger, we require $E_{T}^{\gamma} > 27$~GeV and $0.2 < |\eta_{\gamma}| < 0.6$ in both data and MC simulation. To decrease the contribution from dijet production, where a jet mimics the photon selection, we require the energy in the calorimeter and momentum in the tracking system contained within a cone of $R=0.4$ around the photon to be less than 1 GeV and 2 \GeVc, respectively. As in the $Z$-jet balancing sample, we require events to have exactly one measured jet with $E_{T} > 15$~GeV and no other jets with (uncorrected) $E_{T} > 3$~GeV within $|\eta| < 2.4$. We also demand the $\Delta\phi$ between the jet and photon to be larger than $3.0$~radians. We further reduce contamination by vetoing events with more than one reconstructed interaction point, and by removing events with $\mett / E_{T}^{\gamma} > 0.8$, which likely contain activity from cosmic rays.

\subsection{Determination of corrections}
\label{sec:jesQG_correction}

We derive separate corrections for the quark- and gluon-jet energy scales in data and simulation using the $Z$-jet and $\gamma$-jet balancing samples in the following way. The balances of the $Z$-jet and $\gamma$-jet systems ($K_{Z}$ and $K_{\gamma}$, respectively) can be expressed as linear combinations of independent quark- and gluon-balance variables ($K_{q}$ and $K_{g}$, respectively), weighted by the sample-specific quark and gluon fractions ($F^{q,g}_{Z, \gamma}$)
\begin{equation}
K_{Z} = F_{Z}^{q}K_{q} + F_{Z}^{g}K_{g} = F_{Z}^{q}K_{q} + (1 - F_{Z}^{q})K_{g}~\text{,}
\label{eq:balance_Z}
\end{equation}
\begin{equation}
K_{\gamma} = F_{\gamma}^{q}K_{q} + F_{\gamma}^{g}K_{g} = F_{\gamma}^{q}K_{q} + (1 - F_{\gamma}^{q})K_{g}~\text{.} \label{eq:balance_gamma}
\end{equation}
Rewriting these expressions by solving for $K_{q}$ and $K_{g}$, we find
\begin{equation}
K_{q} = \frac{1}{F_{\gamma}^{q} - F_{Z}^{q}} [ (1 - F_{Z}^{q})K_{\gamma} - (1 - F_{\gamma}^{q})K_{Z} ]~\text{,}
\label{eq:balance_q}
\end{equation}
\begin{equation}
K_{g} = \frac{1}{F_{\gamma}^{q} - F_{Z}^{q}} [ F_{\gamma}^{q}K_{Z} - F_{Z}^{q}K_{\gamma} ]~\text{.}
\label{eq:balance_g}
\end{equation}
These expressions apply separately to experimental data and simulated data, yielding a different balance in data and Monte Carlo simulation ($K\textsuperscript{data}$ and $K\textsuperscript{MC}$, respectively) and may include a dependence on the energy of the jet, with $F_{Z, \gamma}^{q} = F_{Z, \gamma}^{q}(E_{T}\textsuperscript{jet})$ and consequently $K = K(E_{T}\textsuperscript{jet})$.

In order to solve for $K_{q}$ and $K_{g}$, we require knowledge of $K_{Z, \gamma}$ and $F_{Z, \gamma}^{q}$. We extract a value of $K_{Z, \gamma}$ as a function of $E_{T}\textsuperscript{jet}$ by constructing the balancing distribution, as defined in Eq.~(\ref{eq:balance_def}), in ranges of $E_{T}\textsuperscript{jet}$, and fit the core of the distribution with a Gaussian shape. We perform these fits separately in data and simulation and use the mean and uncertainty on the mean of the fitted Gaussian shape as the value of $K_{Z, \gamma}(E_{T}\textsuperscript{jet})$ and its uncertainty. We use this estimation of the most probable value in order to avoid effects from a small fraction of highly mismeasured jets, which may strongly bias the mean and median of the distribution.

\begin{figure*}[htbp]
\centering
\subfigure[]{%
  \includegraphics[width=0.45\textwidth]{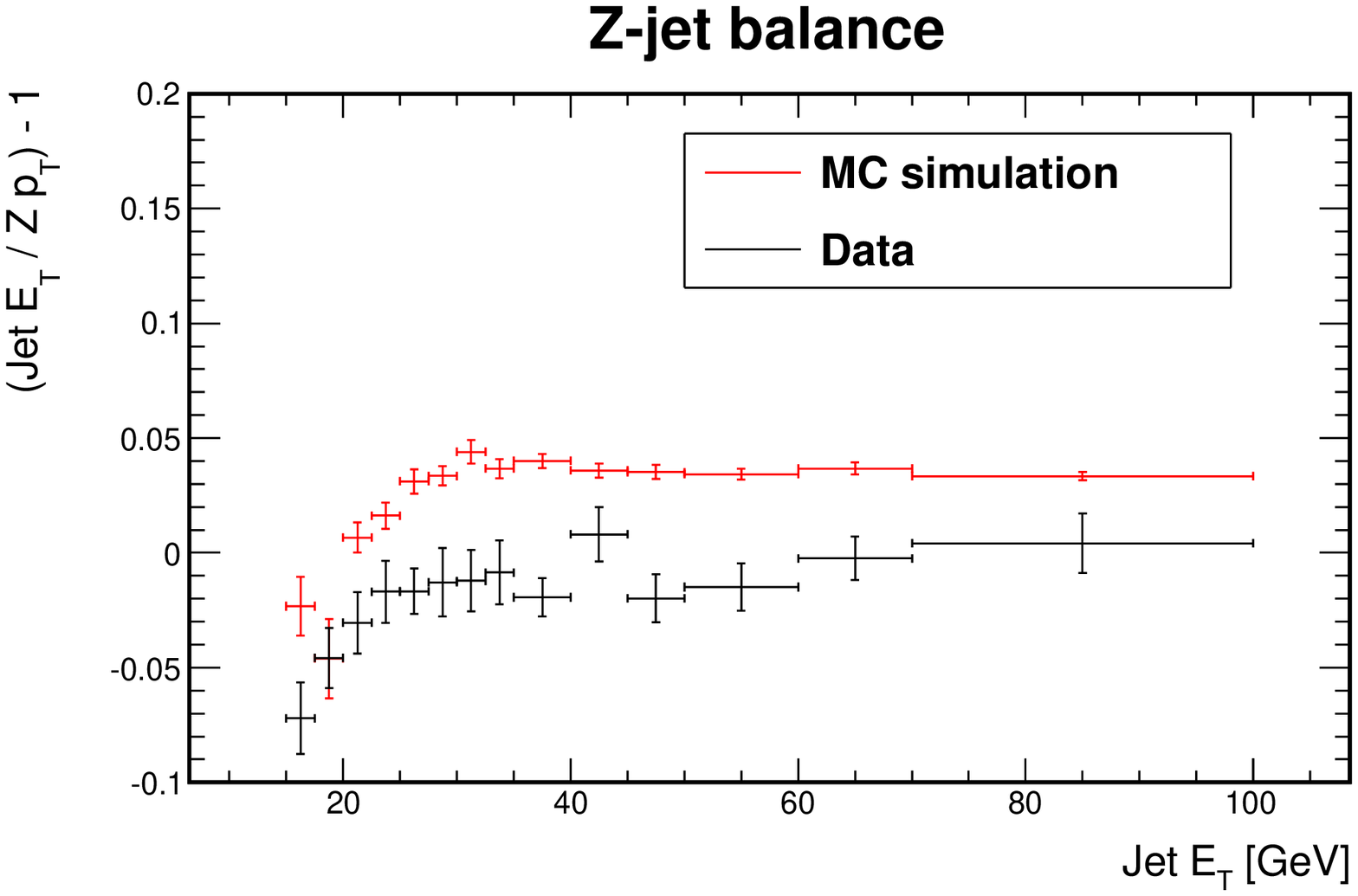}
  \label{fig:balance_z}}
\subfigure[]{%
  \includegraphics[width=0.45\textwidth]{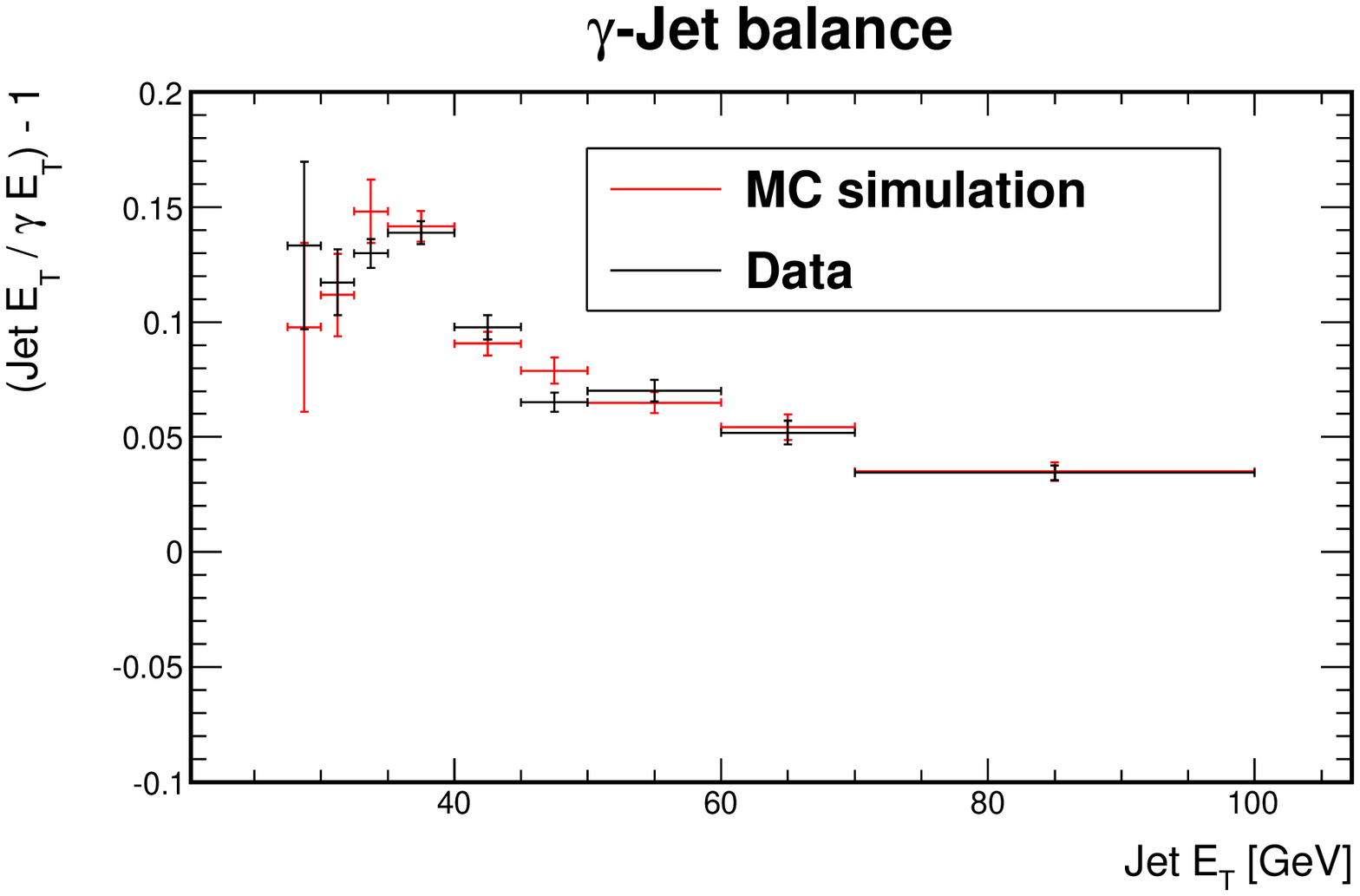}
  \label{fig:balance_gamma}}
\caption{Balancing distributions, \subref{fig:balance_z}~$K_{Z}$ and \subref{fig:balance_gamma}~$K_{\gamma}$, in data and MC simulation as a function of $E_T\textsuperscript{jet}$. The uncertainties include solely the contribution from the fluctuations in the mean of a Gaussian fit to the balancing distributions in bins of $E_T\textsuperscript{jet}$.}
\label{fig:balance_z-gamma}
\end{figure*}

The distributions of $K_{Z}$ and $K_{\gamma}$ in data and simulated data are shown in Fig.~\ref{fig:balance_z-gamma}. Not only are jets measured poorly (the balance does not average to zero), but in the $Z$-jet balancing sample, largely dominated by gluon jets, there is significant disagreement between the correction factors for simulated jets and those in data. We do not see a similar disagreement in the $\gamma$-jet balancing sample, indicating that the simulation models the behavior of the jets in this quark-jet dominated sample accurately.

We determine $F_{Z, \gamma}^{q}$ from simulation by matching jets to their originating partons. In the $\gamma$-jet balancing sample, the quark fraction is about $85\%$ at $E_{T}\textsuperscript{jet} \approx 30$~GeV and drops to about $71\%$ at $E_{T}\textsuperscript{jet} \approx 70$~GeV. In the $Z$-jet balancing sample, these fractions are roughly  $38\%$ and $49\%$, respectively, in the same $E_{T}\textsuperscript{jet}$ regions.
In data, it is not possible to directly match jets to their originating parton, and we must therefore rely on simulation to extract values of $F_{Z, \gamma}^{q}(E_{T}\textsuperscript{jet})$. Because we are trying to correct for discrepancies in the reconstruction of quark and gluon jets between data and simulation, we cannot simply use the simulation-derived $F_{Z, \gamma}^{q}$ values from each jet $E_{T}$ bin. Rather, we parametrize $F_{Z/\gamma}^{q}$ from simulation as a function of $p_{T}^{Z/\gamma}$%:
%$$F_{Z/\gamma}^{q}\textsuperscript{MC}(p_{T}) = a + e^{bp_{t} + c}$$
and determine $F_{Z/\gamma}^{q}\textsuperscript{data}$ in each jet $E_{T}$ bin of the data based on the $p_{T}^{Z}$ or $p_{T}^{\gamma}$ distribution in that bin.

\begin{figure*}[htbp]
\centering
\subfigure[]{%
  \includegraphics[width=0.45\textwidth]{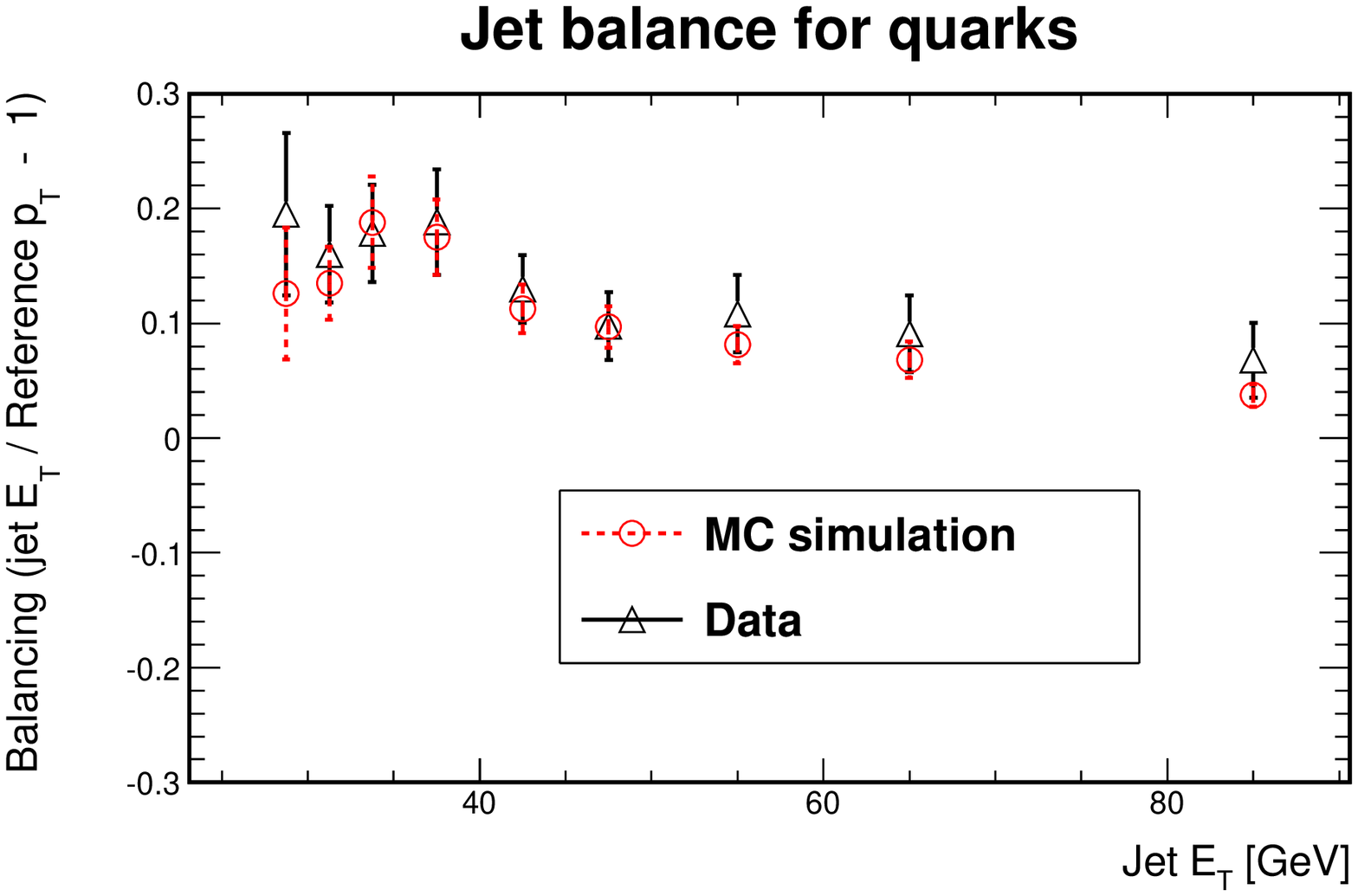}
  \label{fig:balance_q}}
\subfigure[]{%
  \includegraphics[width=0.45\textwidth]{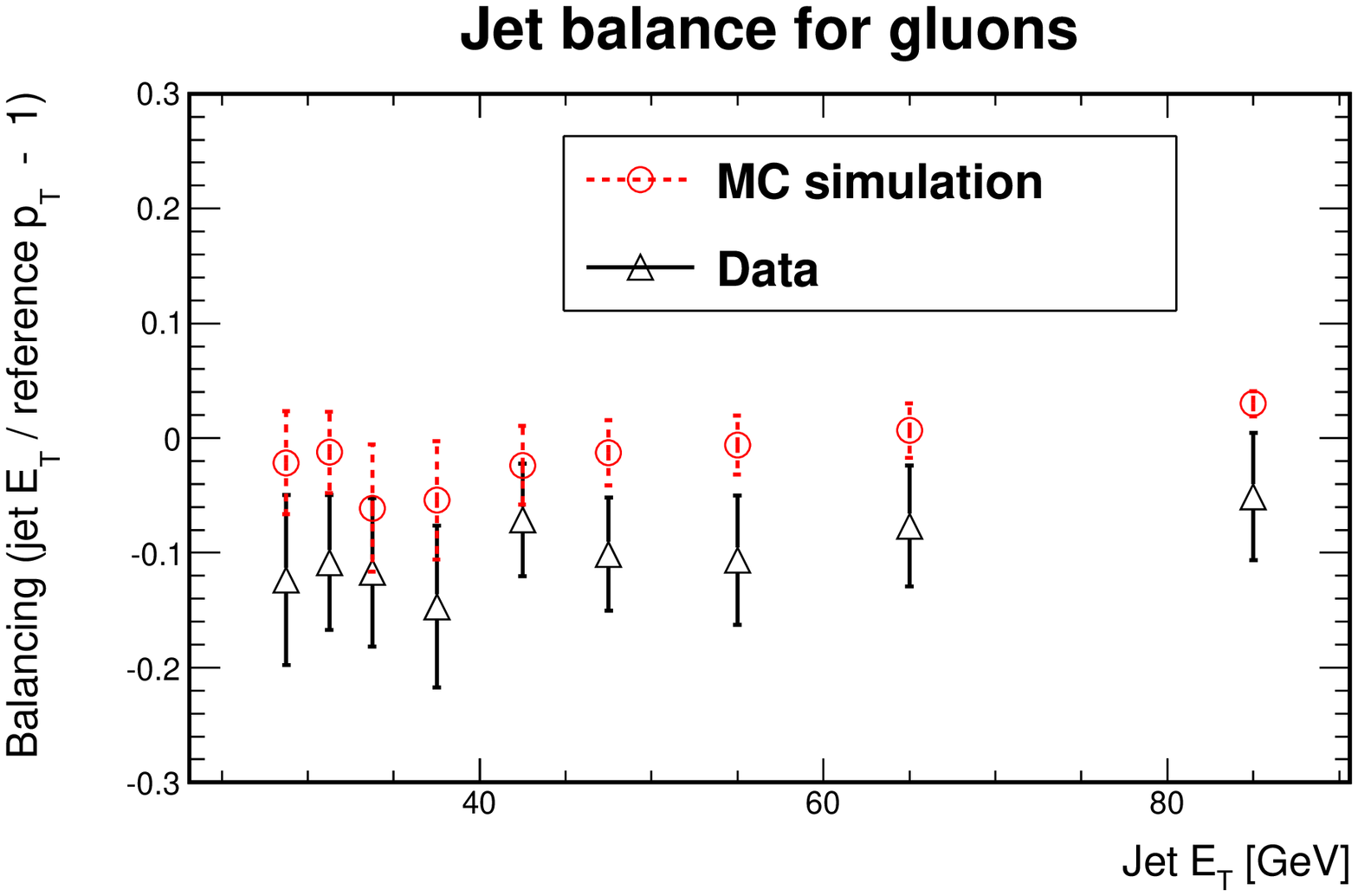}
  \label{fig:balance_g}}
\caption{Derived balancing variable for \subref{fig:balance_q}~quark jets, $K_{q}$, and \subref{fig:balance_g}~gluon jets, $K_{g}$, in data and MC simulation as a function of  $E_T\textsuperscript{jet}$. The uncertainties on each point are from the uncertainties from the mean of the Gaussian fit and the uncertainties on the quark fractions, added in quadrature.}
\label{fig:balance_q-g}
\end{figure*}

Using Eqs.~(\ref{eq:balance_q}-\ref{eq:balance_g}), we construct distributions of $K_{q}$ and $K_{g}$ as functions of the jet $E_{T}$, as shown in Fig.~\ref{fig:balance_q-g}. We see good agreement between data and simulation in $K_{q}$ but poorer agreement in $K_{g}$, where the data correction appears consistently lower than that for simulation. This suggests that the MC simulation is systematically overestimating gluon jet energies, relative to the data.

\begin{figure*}[htbp]
\centering
{\includegraphics[width=0.7\textwidth]{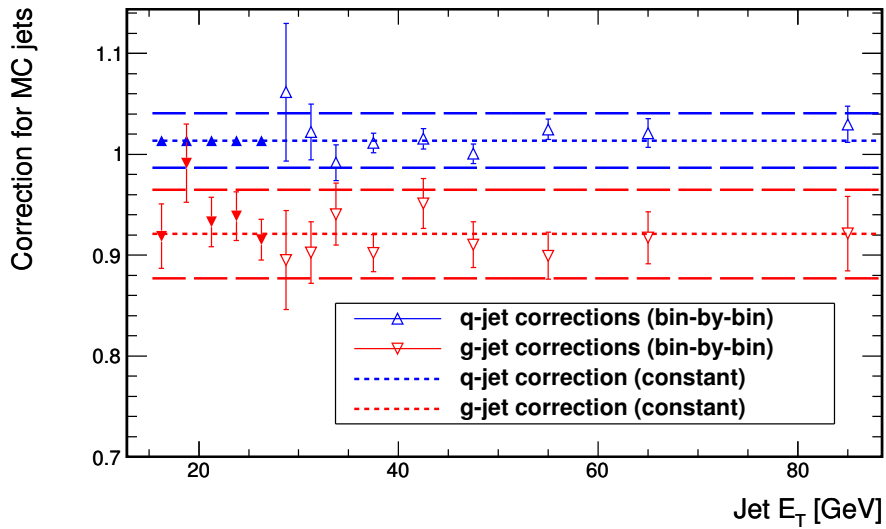}}
%{\includegraphics[width=0.45\textwidth]{PLOTS_zjet_balance_final/balance_TotalSys.eps}}
\caption{Derived correction for simulated quark jets and gluon jets as a function of $E_T\textsuperscript{jet}$. The open triangles represent corrections derived using both $\gamma$-jet and $Z$-jet balancing samples, while the filled triangles represent the assumed uniform correction for quarks and the corresponding correction for gluons calculated from the $Z$-jet balancing sample alone. The uncertainties shown are statistical only. The short dashed lines are the fits of the correction to a constant across jet $E_{T}$ (as opposed to the corrections in each bin of $E_T\textsuperscript{jet}$), and the long dashed lines represent the total systematic uncertainty bands on that constant correction, further described in Sec.~\ref{sec:jesQG_systematics}.}
\label{fig:JES_correction}
\end{figure*}

Using the distributions of $K_{q}$ and $K_{g}$, we determine the corrections that need to be applied to simulated jets in order to best match the energy scale of the data. These MC simulation corrections are defined as $(K_{q}\textsuperscript{data} + 1)/(K_{q}\textsuperscript{MC} + 1)$ for quark jets and $(K_{g}\textsuperscript{data} + 1)/(K_{g}\textsuperscript{MC} + 1)$ for gluon jets, as shown in Fig.~\ref{fig:JES_correction}. Due to the photon trigger used to select the $\gamma$-jet balancing sample, reliable balancing information is not available for jets with transverse energies smaller than 27.5 GeV in that sample, limiting the range of applicability of the corrections. Since we are interested in jets of energies extending down to 20 GeV, we extrapolate the quark-jet energy correction derived for jets with $E_{T} > 27.5$~GeV to lower jet energies and use the $Z$-jet balancing sample to extract a gluon correction assuming this extrapolated quark correction.

The quark and gluon corrections' dependence on jet energy are accurately modeled by a constant for jets with $E_{T} > 15$~GeV. Quark jet energies in simulation should be increased by approximately $1.4\%$ to more accurately match the data, while gluon jet energies should be decreased by approximately $7.9 \%$.

\subsection{Uncertainties on simulated jet energy corrections}
\label{sec:jesQG_systematics}

We consider the following sources of uncertainty on the correction factors.
\begin{enumerate}
\item \textit{Statistical uncertainty}: We use the standard deviation of the corrections to the simulation for each jet $E_{T}$ bin around the assumed uniform correction function. This is an uncertainty of $\pm 2.0\%$ for quark jet energies, and $\mp 2.5 \%$ for gluon jet energies (the uncertainties on quark- and gluon jet energy corrections are anticorrelated).
\item \textit{$F_{Z}^{q}$}: We compare the distribution of a quark-gluon discriminant parameter (described in detail in Sec.~\ref{sec:QGvalue}) in data and simulation, and fit the data distribution using quark and gluon templates from simulation. We take the average deviation of the value determined for this quark-gluon discriminant from the nominal MC simulation value as a systematic uncertainty on $F^{q}$, constant across jet $E_{T}$. This uncertainty is approximately $10\%$. Here we vary the calculated quark fraction in the data $Z$-jet balancing sample by $\pm 10\%$ and recalculate the corrections for quark and gluon jets. This translates to an uncertainty of $\pm 0.6\%$ for quark jet energies and $\mp 2.1\%$ for gluon jet energies.
\item \textit{$F_{\gamma}^{q}$}: We follow a similar procedure of fitting the quark-gluon discriminant parameter in the $\gamma$-jet sample and obtain a similar uncertainty of $\pm 10\%$ on the quark fraction. This translates to an uncertainty of $\pm 1.8\%$ for quark-jet energies, and $\mp 2.7\%$ for gluon jet energies.
\item \textit{Low $E_{T}$ extrapolation}: We check the dependence of the gluon jet energy corrections on the assumed quark jet corrections for low-$E_{T}$ jets by varying the quark jet $E_{T}$ for these jets by $\pm 2\%$. We see a small change in the gluon energy corrections, which translates to $\mp 0.4\%$ of the gluon jet energy. 
\item \textit{Number of interaction vertices dependence}: The $\gamma$-jet balancing sample incorporates a requirement on the number of reconstructed interaction vertices to reduce contamination from pileup. The $Z$-jet balancing sample does not have such a requirement, due to a much smaller background contribution and in order to retain as many events as possible. We check for any bias in the corrections resulting from the effect of this requirement by looking for any shift in the corrections when the requirement is placed on the $Z$-jet balancing sample. We see a change in the quark-jet energies of $\pm0.2\%$, and the gluon jet energies of $\mp1.2\%$.
\end{enumerate}
The uncertainties are summarized in Table~\ref{tab:jesQG_systematics}. Because the corrections shift the energy response in the simulation to more accurately match the data, the quark jet and gluon jet energy correction uncertainties are anticorrelated. The uncertainties are similar in magnitude to the default CDF jet energy scale uncertainties~\cite{cdf_JES}.

\begin{table}[hbt]
\begin{center}
\begin{ruledtabular}
\begin{tabular}{llcc}
& & Quark jets& Gluon jets\\ \hline
%Jet energy & \multirow{2}{*}{} & \multirow{2}{*}{1.014} & \multirow{2}{*}{0.921} \\ %\hline
%correction & & & \\ \hline
Jet energy correction &  & 1.014 & 0.921 \\ \hline
%correction & & & \\ \hline
Uncertainty & Statistical & 0.020 & 0.025 \\
& $F_{Q}^{Z-\text{jet}}$ & 0.006 & 0.021 \\
& $F_{Q}^{\gamma-\text{jet}}$ & 0.018 & 0.027 \\
& Low $E_{T}$ extrapolation & & 0.004 \\
& $N_{vert}$ difference & 0.002 & 0.012 \\ \hline
& Total uncertainty &  $\pm 0.027$ & $\mp 0.044$  \\
\end{tabular}
\end{ruledtabular}
\caption{Summary of the additional jet energy corrections applied to simulated jets and their uncertainties.}
\label{tab:jesQG_systematics}
\end{center}
\end{table}

\section{Artificial Neural Network Quark-to-Gluon Discriminant}
\label{sec:QGvalue}
In this analysis, we search for two high-$p_{T}$ leptons from the decay of a $Z$ boson and two jets from a $W \rightarrow q\bar{q'}$ or $Z \rightarrow q\bar{q}$ decay. Thus, the two signal jets are quark jets. Conversely, the dominant background, two jets produced in association with a $Z \rightarrow \ell^{+}\ell^{-}$ decay, contains a significant fraction of gluon jets (of the order of 50\%). The ability to separate quark jets from gluon jets is therefore useful for increasing sensitivity to \textit{ZW} and \textit{ZZ} production.

For a given energy, gluon jets, due to their higher color charge, tend to feature a higher particle multiplicity and be spatially broader in the detector than light-quark ($u$, $d$, and $s$) jets. We attempt to quantify the spatial spread of jets using a collection of artificial neural-networks (NNs) trained to separate gluon jets from light-flavor quark jets. We refer to the output of the final NN as the jet quark-to-gluon discriminant value (or jet QG value). We calibrate the response of the final NN in MC simulation to match the response in data based on a $W \rightarrow \ell\nu +$ 1 jet event sample. The tagging efficiency and mistag rate associated to a requirement on the jet QG value are obtained from two independent event samples:  $W \rightarrow \ell\nu +$~2 jets events, which are representative of the $Z +$~jets background; and $t\bar{t} \rightarrow b\bar{b} \ell\nu q\bar{q'}$ events, which contain two non-heavy-flavor jets from the hadronic decay of a $W$ boson, similar to the diboson signal.

\subsection{Jet QG definition}
\label{sec:QGvalue_definition}
A total of three NNs contribute to the final QG discriminant. The initial two networks separate quark and gluon jets by exploiting distinctive features in the distribution of energies reconstructed in calorimeter towers and momenta of charged particles reconstructed in the tracking chambers. Thus, every jet is assigned a \textit{tower} NN value and \textit{track} NN value, which are the outputs of these networks. These two NN values are then used as inputs to a third NN.
%Variables sensitive to the spread and collimation of the jets independent of the quark or gluon hypothesis are also included in that third NN.

Each of the NNs is trained using simulated samples of jets matched to either a light-flavor-quark or gluon with $p_{T} > 20$~\GeVc~ within $\Delta R = 0.4$ of the center of the jet and further requiring that no additional partons with transverse momenta exceeding 8 \GeVc~are present within $\Delta R = 0.7$. These jets are selected from a $Z \rightarrow \mu^{+}\mu^{-} + 2$ parton {\sc alpgen} sample, interfaced with {\sc pythia} showering. Each NN is a feed-forward multilayer perceptron with a hyperbolic-tangent-like response function~\cite{tmva}. The networks are trained on $100\, 000$ quark and gluon jets and tested for biases in overtraining on samples containing $500\, 000$ quark and gluon jets. Gluon-jet distributions are reweighted to match the $E_{T}$ and $\eta$ distributions of the quark jets to remove any discrimination power coming solely from these variables.

\begin{figure*}[htbp]
\centering
\subfigure[]{%
  \includegraphics[width=0.45\textwidth]{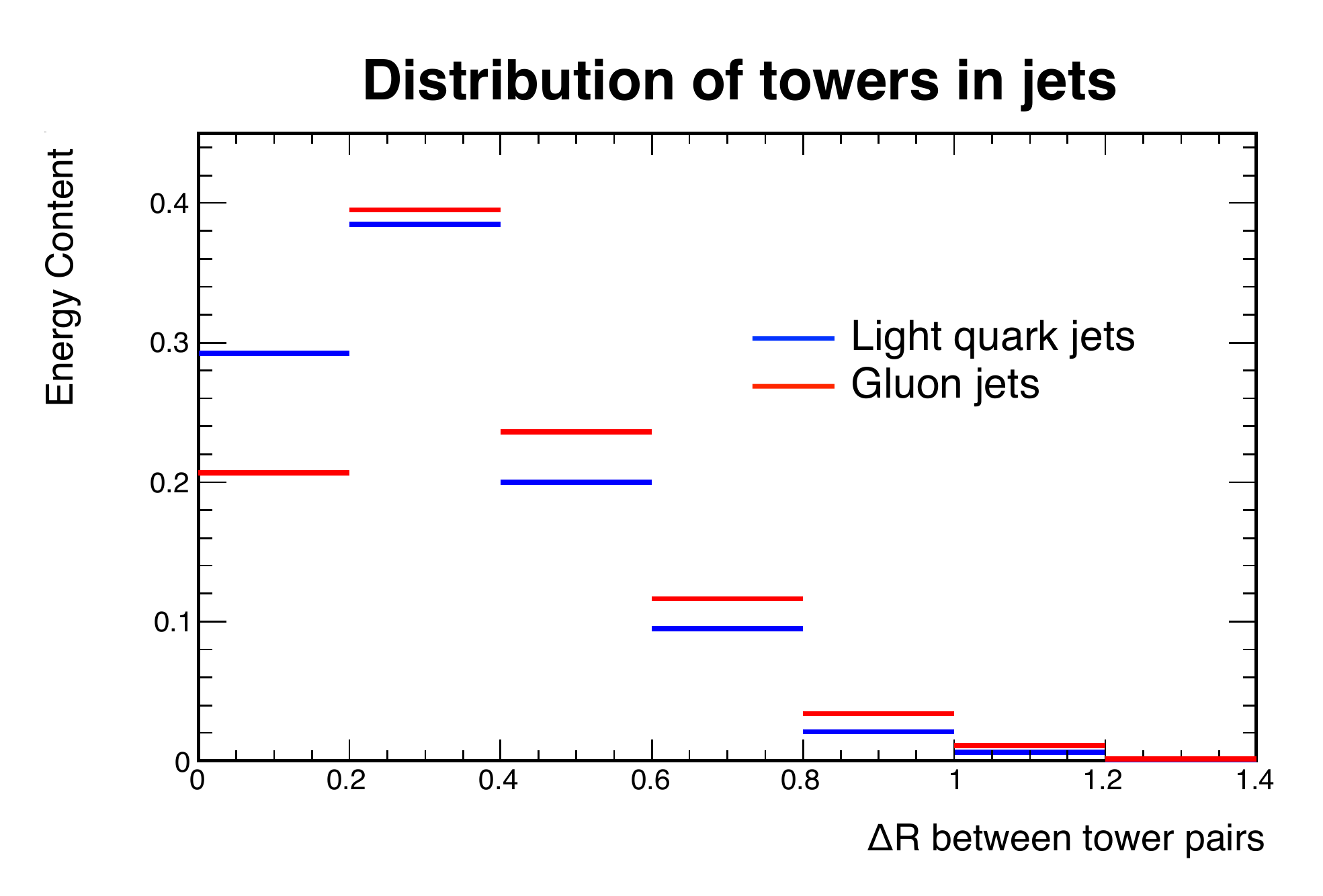}
  \label{fig:nn_inputs_tower}}
\subfigure[]{%
  \includegraphics[width=0.45\textwidth]{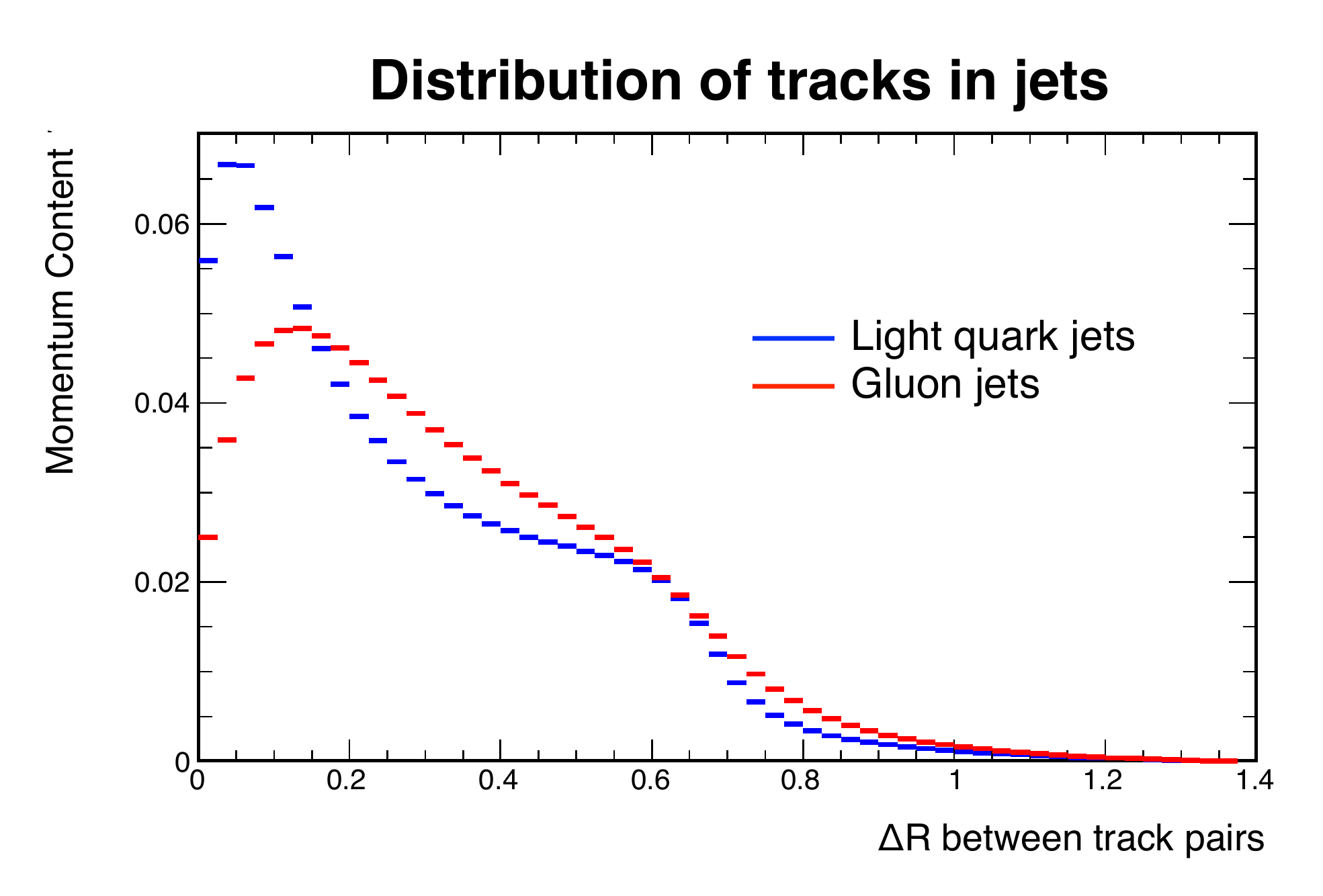}
  \label{fig:nn_inputs_track}}
\caption{Typical distributions of energy (momentum) content of jets as a function of the $\Delta R$ \subref{fig:nn_inputs_tower}~between pairs of towers and \subref{fig:nn_inputs_track}~between pairs of tracks in light-flavor quark and gluon jets in simulation. }
\label{fig:nn_inputs}
\end{figure*}

\begin{figure*}[htbp]
\centering
\subfigure[]{%
  \includegraphics[width=0.45\textwidth]{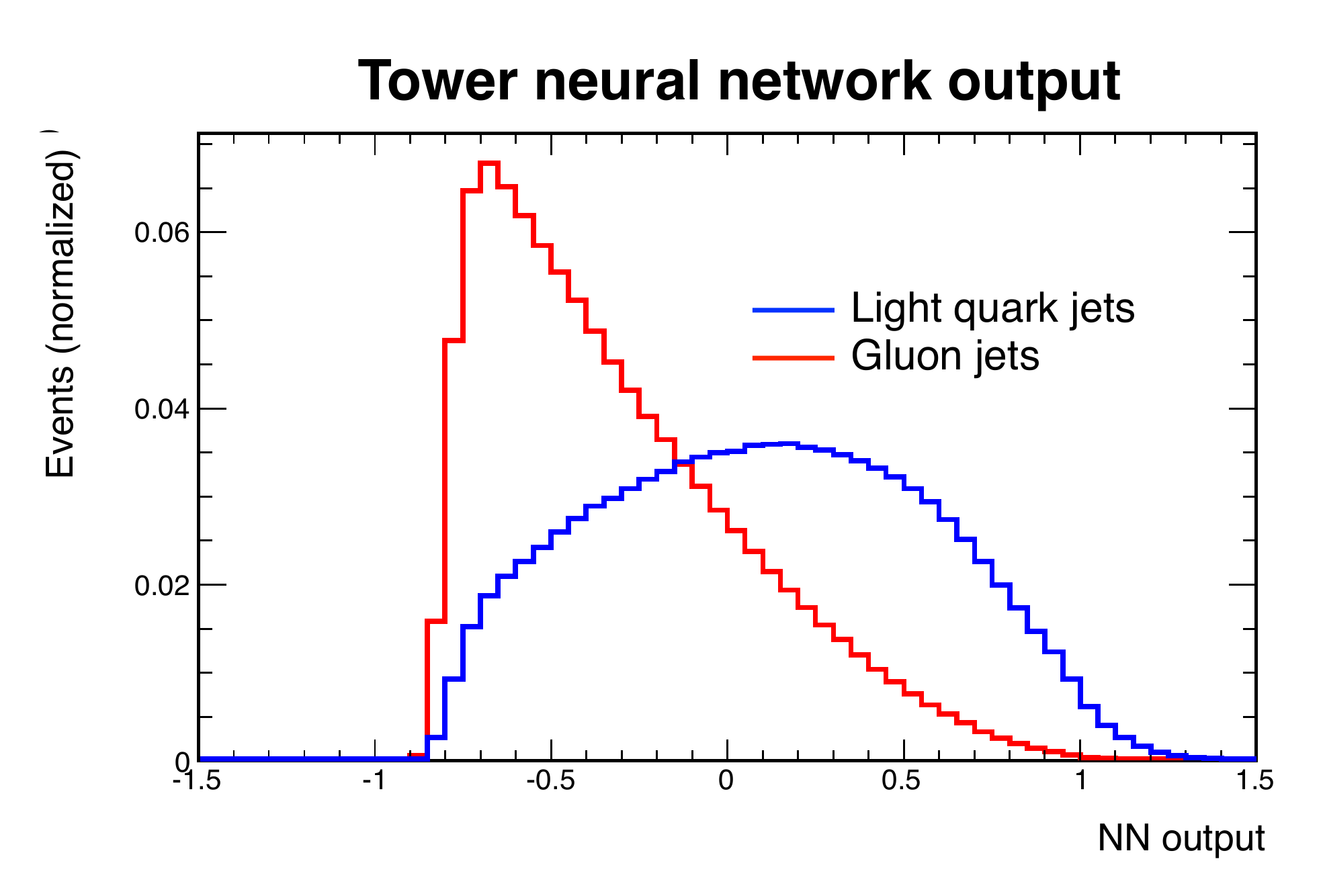}
  \label{fig:nn_output_tower}}
\subfigure[]{%
  \includegraphics[width=0.45\textwidth]{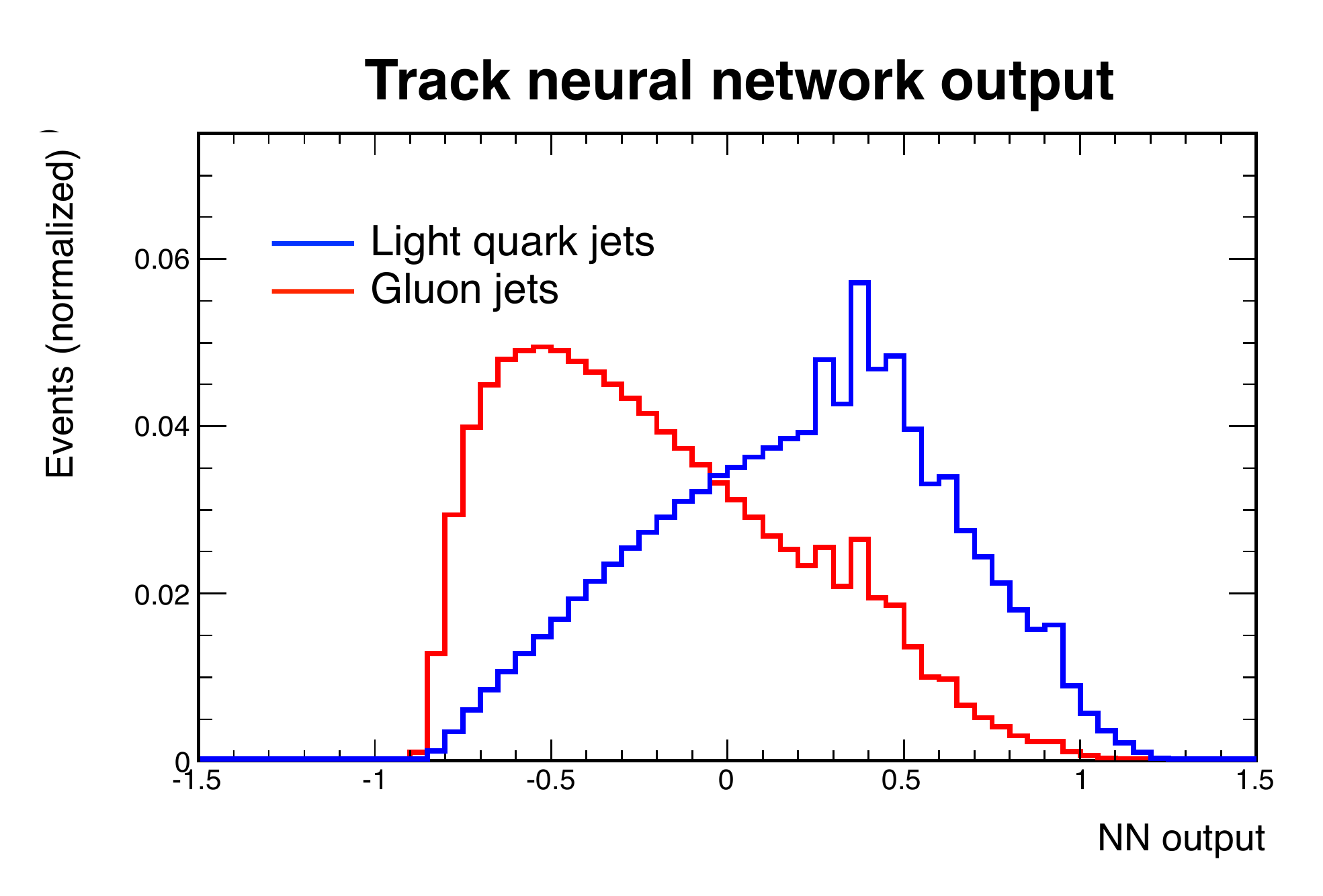}
  \label{fig:nn_output_track}}
\caption{Distributions of the outputs of the NNs processing ~\subref{fig:nn_output_tower}~tower information and \subref{fig:nn_output_track}~track information in light-flavor quark and gluon jets in simulation. Higher NN scores indicate jets that are more quark-like.}
\label{fig:nn_output}
\end{figure*}

For each jet we obtain a list of the calorimeter towers within a cone of $\Delta R = 0.7$. Each tower has a location coordinate, $(\eta,\phi)$, and energy deposition $E$ associated with it. We construct a distribution of the distance, $\Delta R$, between all pairs of towers within the jet and weight each tower pair by its relevance in terms of energy to obtain a distribution that characterizes the spatial spread of the energy within each jet. The weight is given by
$$\frac{E_{i}E_{j}}{0.5\left[(\Sigma E)^{2} - \Sigma E^{2}\right]}~\text{,}$$
where $E_{i}$ and $E_{j}$ are the energies detected in the two towers of the pair, $\Sigma E$ is the sum of the energy in all towers within a cone of $\Delta R = 0.7$ around the jet, and $\Sigma E^{2}$ is the sum of the square of the energies of each tower in that same cone. The denominator is chosen to normalize the sum of the weights of all tower pairs to unity. We sample this distribution in 56 intervals (bins) of size $\Delta R_\text{bin} = 0.025$ for $0.0 < \Delta R < 1.4$, where the contents of the first three bins are empty due to the segmentation of the calorimeter. Typical distributions of the weighted $\Delta R$ between tower pairs for quark and gluon jets are shown in Fig.~\ref{fig:nn_inputs}, using a larger bin size. The outputs of the tower NN for quark and gluon jets using the training and testing samples are shown in Fig.~\ref{fig:nn_output}.

We follow a similar prescription using tracks within a cone of $\Delta R = 0.7$ around each jet, using the tracks' locations in $(\eta,\phi)$ (with respect to the primary vertex) and momenta $p$ to obtain a distribution of the distance between pairs of tracks (in $\Delta R$), with each pair weighted by the momentum carried by that pair, or 
$$\frac{p_{i}p_{j}}{0.5\left[(\Sigma p)^{2} - \Sigma p^{2}\right]}~\text{,}$$
where $p_{i}$ and $p_{j}$ are the magnitude of the momenta of the charged particles in the pair, $\Sigma p$ is the scalar sum of the momenta carried by all charged particles within a cone of $\Delta R = 0.7$ around the jet, and $\Sigma p^{2}$ is the sum of the square of the momenta of each charged particle within that same cone. We require all contributing charged particles to come from the primary vertex and have $p_{T} > 0.4$~\GeVc. We split the $\Delta R$ between track pairs distribution into the same 56 intervals (bins) as used in the tower NN, and the content of each bin is used as an input into the track NN.

Typical distributions of $\Delta R$ between track pairs for quark and gluon jets are shown in Fig.~\ref{fig:nn_inputs}. Light-flavor quark jets tend to peak at low $\Delta R$, indicating that they are rather collimated, while gluon jets tend to have a higher mean-valued $\Delta R$ distribution. The bin contents of these $\Delta R$ distributions are used as inputs into NNs that discriminate between quark and gluon jets.

The outputs of the track NN for quark and gluon jets using the training and testing samples are shown in Fig.~\ref{fig:nn_output}. Higher NN scores indicate jets that are more quark-like. We see good performance in both the tower and track NNs. The cusps in the track NN distribution are associated to jets containing only two charged-particle tracks located inside a cone of $\Delta R = 0.7$, and thus have only one nonzero bin in their distributions of $\Delta R$ between track pairs.

\begin{figure*}[htbp]
\centering
{\includegraphics[width=0.7\textwidth]{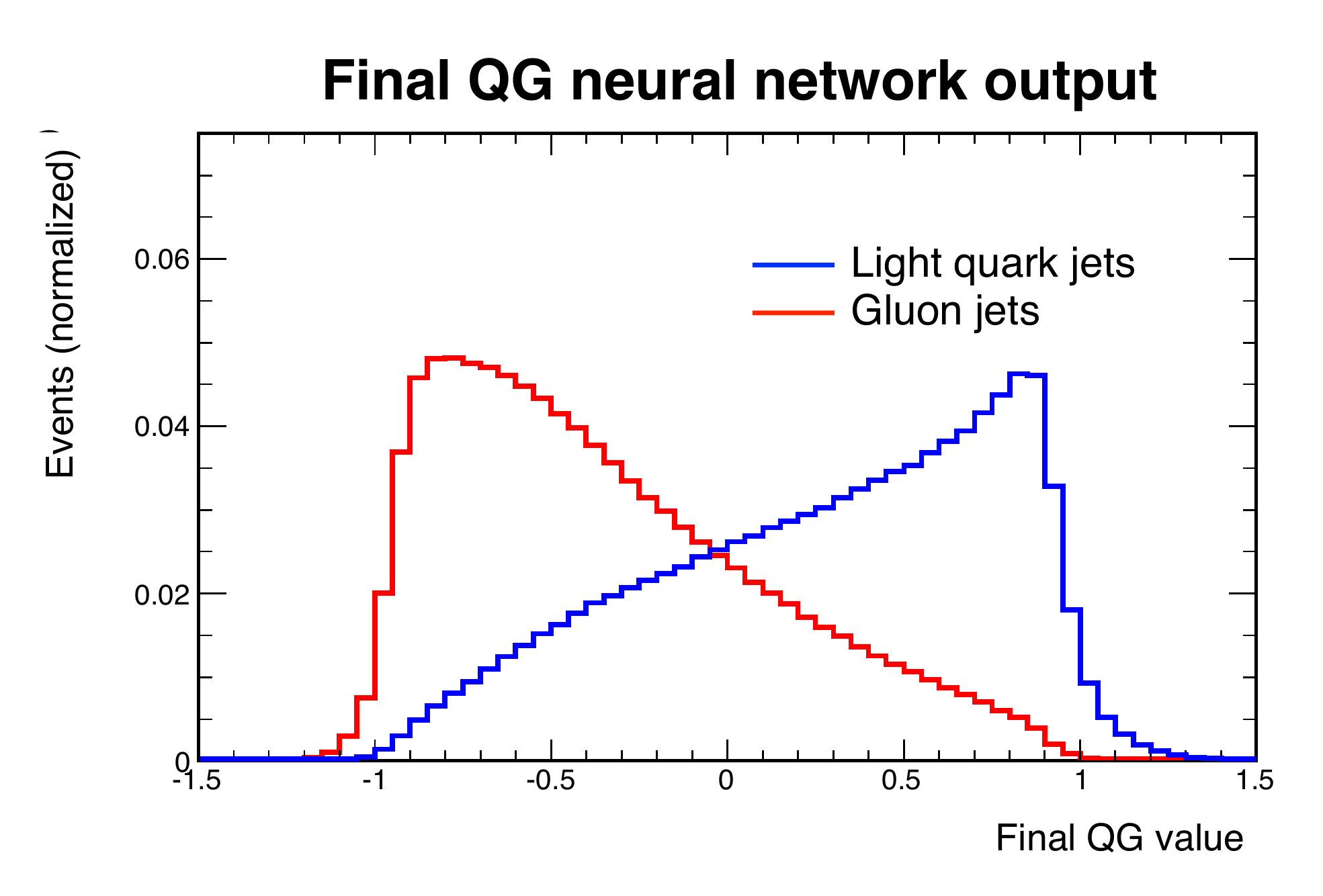}}
%{\includegraphics[width=0.45\textwidth]{figures_QG/FinalQG_NN_bless.eps}}
\caption{Distribution of the output of the final NN for light-flavor-quark and gluon jets in simulation. Higher NN scores indicate jets that are more quark-like.}
\label{fig:nn_qg}
\end{figure*}

The final NN uses both the tower and track NN values as inputs, along with other jet variables that provide discrimination power between quark jets and gluons: the ratio of $\Sigma E$ associated to towers within a cone of $\Delta R=0.4$ to $\Sigma E$ associated to towers within a cone of $\Delta R=0.7$; the ratio of $\Sigma p$ associated to charged particles within a cone of $\Delta R=0.4$ to $\Sigma p$ associated to charged particles within a cone of $\Delta R=0.7$; the number of contributing towers with nonzero energy in cones of $\Delta R=0.4$ and $0.7$; the number of contributing charged-particle tracks in cones of $\Delta R=0.4$ and $0.7$; and the jet EM fraction. Additionally, other variables that affect the shape of the $\Delta R$ distributions, independent of whether the jet originates from a quark or gluon, are included: the jet $E_{T}$; the jet $\eta$; and, the number of reconstructed interaction vertices in the event. The output of this final NN is shown in Fig.~\ref{fig:nn_qg} for light-flavor-quark and gluon jets using the training and testing samples. In simulated jets, we see significant separation between quark and gluon jets using this discriminant.

\subsection{Jet QG calibration and performance}
\label{sec:QGvalue_calibration}
The response of the NN quark-to-gluon discriminant may differ between data and MC simulation, especially since uncorrected tower energies are used in the construction of the tower NN. Since the signal and most backgrounds are modeled with simulated data, we calibrate the simulation response to match the response in data. We use a control region of independent events with a jet composition similar to that of the final state, $W \rightarrow \ell\nu +$ 1 jet events. We then validate the calibration and establish uncertainties on the modeling using control samples of data with features similar to the signal and samples enriched in the dominant backgrounds: $t\bar{t}$ decays in lepton+jets final states and $W \rightarrow \ell\nu+2$ jet events, respectively.

To form the $W + 1$ jet calibration sample, we choose data collected with the standard high-$E_{T}$ ($p_{T}$) central electron (muon) triggers and select events with exactly one central ($|\eta|<1.0$) electron (muon) with $E_{T}$ ($p_{T}$) $> 20$~\GeVc. To select events consistent with a $W \rightarrow \ell\nu$ decay, we also require a significant missing transverse energy, $\mett > 25$~GeV, and a reconstructed transverse mass~\cite{mTdef} consistent with leptonic $W$ boson decays, $m_{T} > 25$~\GeVcc. To further suppress any contributions from multijet events where a jet mimics the lepton $+ \mett$ signature, we require that the $\mett$ is not aligned with any reconstructed jet ($\Delta\phi(\mett,$jet$) > 0.2$~radians) and that the $\mett$-significance---a dimensionless quantity comparing the observed $\mett$ against the energy resolution of jets, soft unclustered particles, and the event topology (see Ref.~\cite{metjj_prl})---be larger than four (one) for events with electrons (muons). We also require that the events in this calibration sample have exactly one jet with $E_{T} > 20$~GeV and $|\eta| < 2.0$.

We consider various processes that contribute to this sample, listed in Table~\ref{tab:qgStudies_events}, and model them using a combination of the {\sc pythia}, {\sc alpgen}, and {\sc madgraph}~\cite{madgraph} event generators interfaced with {\sc pythia} for showering. The dominant contribution is $W \rightarrow \ell\nu$ production in association with one jet, which is modeled using {\sc algpen}. As we are largely concerned with the agreement in shapes between data and simulation, we scale the simulation distributions to match the data. Additionally, we reweight the simulation to match the jet $E_{T}$ and $\eta$ distributions in data to remove these variables as a possible causes for mismodeling of the jet QG value.

\begin{table*}[!hbt]
\begin{minipage}[b]{\textwidth}\centering
\begin{ruledtabular}
\begin{tabular}{lcc}
& $W$+jets selection & $t\bar{t}$ selection\\ \hline
$W$+jets & $21\,500 \pm 2\,200$ & $38.7 \pm 3.9$ \\
$W$+$b$ jets & $940 \pm 380$ & $13.8 \pm 5.5$ \\
$Z$+jets & $1\,250 \pm 130$ & $3.1 \pm 0.3$ \\
$Z$+$b$ jets & $86 \pm 34$ & $1.4 \pm 0.6$ \\
$WW + WZ$ & $1\,386 \pm 83$ & $5.9 \pm 0.4$\\
Single top-quark & $767 \pm 77$ & $19.6 \pm 2.0$\\
$t\bar{t}$ & $1\,378 \pm 83$ & $469 \pm 28$\\
    $t\bar{t}$ ($b$ jets) & & $108 \pm 7$\\
    $t\bar{t}$ ($q$ jets) & & $361 \pm 22$\\ \hline
Total expected & $27\,300 \pm~2\,200$ & $551 \pm 30$ \\
Data & $27\,319$ & 579 \\
\end{tabular}
\end{ruledtabular}
%\footnotetext[1]{}
\end{minipage}
\caption[Number of events in $W+$~jets and $t\bar{t}$ selections used for QG value efficiency/mistag rate studies]{Number of events in the $W$+2 jets and $t\bar{t}$ lepton+jets region, showing only the uncertainties assigned on the normalization of each sample. The $W +$~jets samples are rescaled to match data in the number of events observed after the $W+$~jets selection. The distinction between $b$ and $q$ jets in the $t\bar{t}$ sample refers to the lower two $b$ness jets: events where both jets are matched to non-$b$-quark jets are labeled \textit{$q$ jets}, while if one of the jets is matched to a $b$ jet, the event is labeled \textit{$b$ jets}.}
\label{tab:qgStudies_events}
\end{table*}

We observe poor modeling of the tower NN values, where the jets in data appear more gluon-like than those in simulated events. The fact that jets in data appear more spatially spread than jets in simulation is consistent with the observed differences in jet energy scales for data and simulation, described in Sec.~\ref{sec:jesQG}: the fraction of the jet energy contained within a cone of $\Delta R = 0.4$ is higher in simulated gluon jets than in gluon jets in data. We correct for these discrepancies by applying a linear shift to the tower NN values observed in simulation in order to match with data using the $W + 1$ jet sample. We apply different linear shifts for jets in the central and plug calorimeters, and for jets in events with different levels of pileup. We apply further corrections to the response of the final NN to more accurately match the correlations of these calibrated tower-NN values with other jet quantities: the number of towers in the jets and the ratio of $\Sigma E$ in a cone of $\Delta R=0.4$ to $\Sigma E$ in a cone of $\Delta R=0.7$. The modeling is more accurate in the track NN than in the tower NN, though we still introduce a similar linear shift in simulated track NN values to more accurately match data. The calibrated variables are input directly into the final NN, without retraining the network.

We further validate the response of the jet QG value by comparing data and MC simulation in a $W \rightarrow \ell\nu+2$ jets event sample and in an event sample dominated by $t\bar{t}$ production where two quark jets originate from the hadronic decay of a $W$ boson. Table~\ref{tab:qgStudies_cuts} summarizes the requirements used to select these two samples: the $W+2$ jet sample is similar to the previously described $W + 1$ jet sample, except for modified jet selections to match those used in the signal region of the $ZV \rightarrow \ell\ell jj$~search. The $t\bar{t}$ selection eschews the $\mett$-significance and $m_{T}$ requirements, used to reduce multijet backgrounds, in favor of requirements on the minimum scalar sum of transverse quantities (jets' $E_{T}$, $\mett$, and the charged lepton's $p_{T}$) in the event, which is effective in removing both multijet and $W + $ jets backgrounds. Because we are interested in selecting the two jets in the $t\bar{t}$ candidate events that come from the decay of a $W$ boson, as opposed to the $b$ jets produced in the $t \rightarrow Wb$ decays, we make use of the jet-$b$ness tagger~\cite{bness_nim}. This multivariate $b$ jet identification algorithm exploits properties of individual charged-particle tracks within a jet, looking at properties characteristic of charged particles originating from $B$-hadron decays. The final score, the output of a NN discriminant that ranges between $-1$ and $1$, is called the \textit{jet $b$ness}, where higher scores identify jets that are more likely to originate from $B$-hadron decays.  We classify the two jets with the highest $b$ness scores in the event as the two $b$ jets, and the remaining two jets as those resulting from a $W \rightarrow q\bar{q'}$ decay.

\begin{table*}[hbt]
\begin{center}
\begin{ruledtabular}
\begin{tabular}{ccc} 
$W$+jets selection &  & $t\bar{t}$ selection\\ \hline
&Central $e$ or $\mu$, $p_{T} > 20$~\GeVc & \\ 
& $\mett > 25$~GeV & \\
$\Delta\phi(\mett,$~nearest jet$)> 0.4$~rad & & $\Delta\phi(\mett,$~nearest jet$)> 0.2$~rad \\
$\mett$-sig $> 4$ ($e$ only) & & \\
$m_{T}(W) > 25$~\GeVcc ($e$ only) & & \\
& & Sum $E_{T} > 300$~GeV \\
$N_{jets}(E_{T} > 20$ GeV$) = 2$ & & $N_{jets}(E_{T} > 20$ GeV$) = 4$ \\
& & 2nd highest $b$ness jet $b$ness $>$ -0.5 \\
1st/2nd jet $E_{T} > 25$~GeV & & 2nd highest $b$ness jets $E_{T} > 20$~GeV\\ 
& & 2 lowest $b$ness jets $E_{T} > 25$~GeV\\ 
& Jet $|\eta| < 2.0$ & \\ 
& $\Delta R$ between jets $> 0.7$ & \\
\end{tabular}
\end{ruledtabular}
\caption{Summary of event selection requirements for the $t\bar{t}$ lepton+jets selection and the $W$+2 jets selection, used to understand the modeling of events in the QG discriminant. Requirements in the center are shared requirements in the two samples.}
\label{tab:qgStudies_cuts}
\end{center}
\end{table*}

\begin{figure*}[htbp]
\centering
%{\includegraphics[width=0.45\textwidth]{figures/w2jet_max_qg_crop.pdf}}
%{\includegraphics[width=0.45\textwidth]{figures/w2jet_min_qg_crop.pdf}}
%{\includegraphics[width=0.45\textwidth]{figures/ttbar_max_qg_crop.pdf}}
%{\includegraphics[width=0.45\textwidth]{figures/ttbar_min_qg_crop.pdf}}
\subfigure[]{%
\includegraphics[trim=0cm 0cm 5cm 0cm, clip=true, width=0.3625\textwidth]{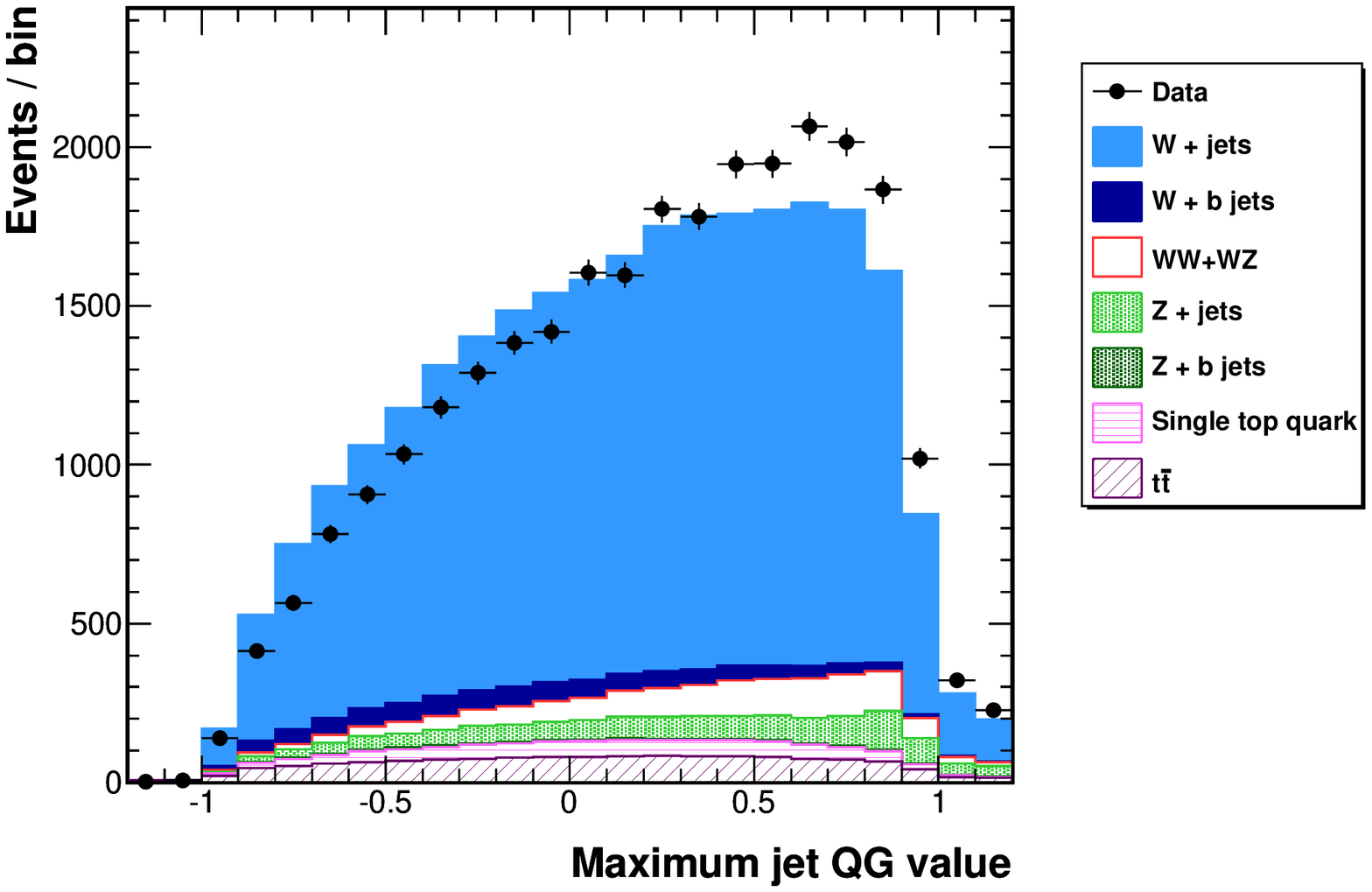}
\label{fig:qgStudies_wjet_max}}
\subfigure[]{%
\includegraphics[trim=0cm 0cm 5cm 0cm, clip=true, width=0.3625\textwidth]{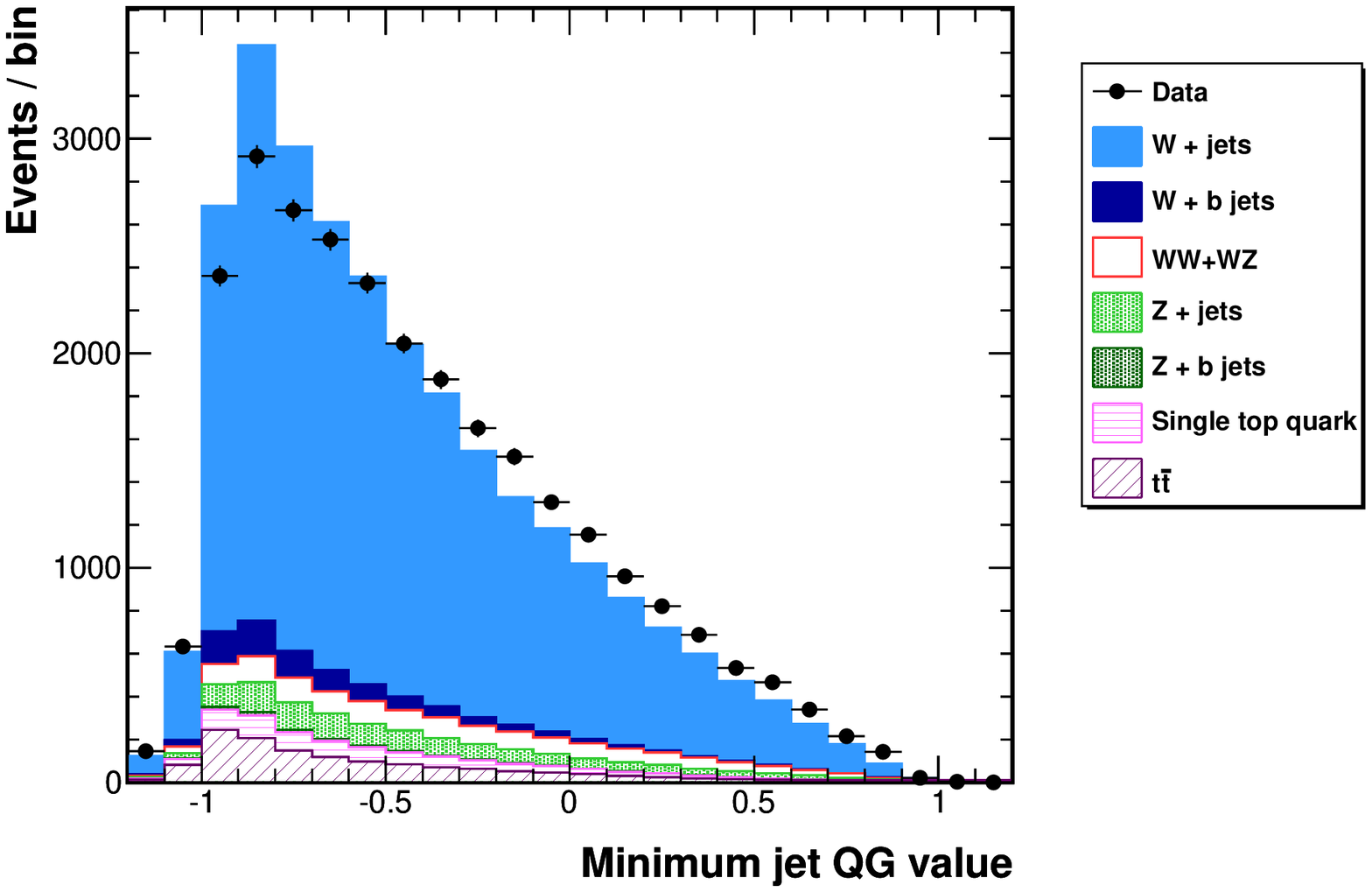}
\label{fig:qgStudies_wjet_min}}
\subfigure{%
\includegraphics[trim=15cm 4cm 0cm 1cm, clip=true, width=0.175\textwidth]{figures_PRD/w2jet_maxQG_final_update_BW.eps}}
\caption{Distribution of \subref{fig:qgStudies_wjet_max}~the maximum and \subref{fig:qgStudies_wjet_min}~minimum jet QG values of the two jets in the $W+2$ jet sample.}
\label{fig:qgStudies_dists_w2jet}
\end{figure*}

\begin{figure*}[htbp]
\centering
\subfigure[]{%
\includegraphics[trim=0cm 0cm 5cm 0cm, clip=true, width=0.3625\textwidth]{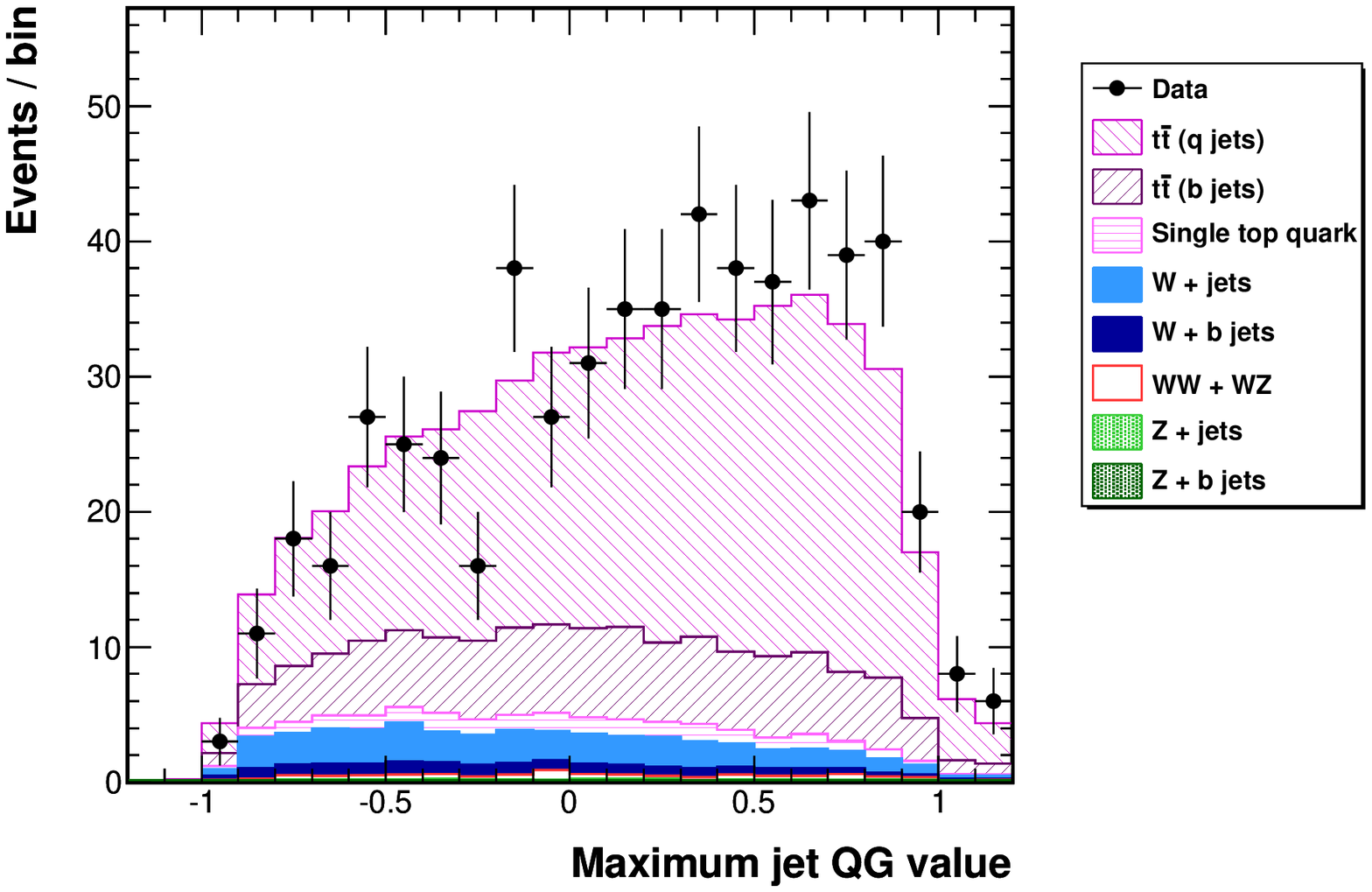}
\label{fig:qgStudies_ttbar_max}}
\subfigure[]{%
\includegraphics[trim=0cm 0cm 5cm 0cm, clip=true, width=0.3625\textwidth]{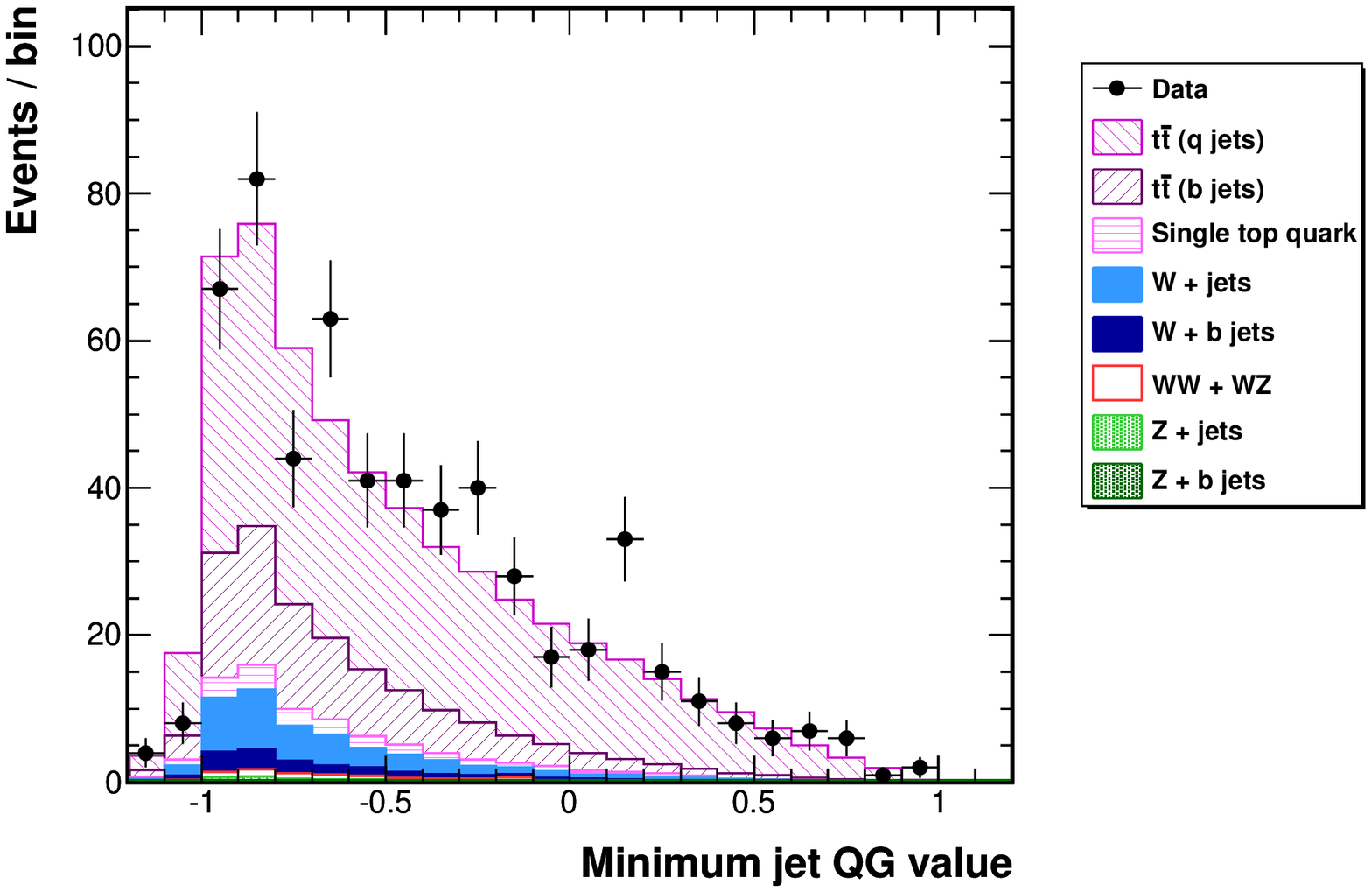}
\label{fig:qgStudies_ttbar_min}}
\subfigure{%
\includegraphics[trim=15cm 4cm 0cm 1cm, clip=true, width=0.175\textwidth]{figures_PRD/ttbar_maxQG_final_update_BW.eps}}
\caption{Distribution of \subref{fig:qgStudies_ttbar_max}~the maximum and \subref{fig:qgStudies_ttbar_min}~minimum jet QG values of the two jets in the $t\bar{t}$ sample. The distinction between $q$ and $b$ jets refers to the lower two $b$ness jets: events where both jets are matched to non-$b$ quark jets are labeled \textit{$q$ jets}, while if one of the jets is matched to a $b$ jet, the event is labeled \textit{$b$ jets}.}
\label{fig:qgStudies_dists_ttbar}
\end{figure*}

Because we are looking for jet QG shape differences between data and simulation that induce acceptance uncertainties when a requirement on the jet QG value is applied, we scale the number of $W+$jet events in simulation to match the yield observed in the $W+2$ jets data. The number of events in each sample is shown in Table~\ref{tab:qgStudies_events}. The distributions of the maximum and minimum QG values of the two jets are shown in Figs.~\ref{fig:qgStudies_dists_w2jet}-\ref{fig:qgStudies_dists_ttbar}. We see fairly good modeling in the $t\bar{t}$ sample, but poorer modeling in the $W+2$ jet sample where, even after calibrations, the jets in simulation appear more gluon-like than the jets in data. We account for this remaining discrepancy between data and simulation below.

We enhance the sensitivity to the signal when forming a \textit{light-flavor-tagged} channel where the minimum jet QG value is greater than $0.0$. We determine a probability for a quark jet to meet this requirement (efficiency), and for a gluon jet to be misidentified as a quark jet (mistag rate), with the $t\bar{t}$ and $W+2$ jet samples. The efficiency measured in data, $e_{D}$ is a function of the QG requirement, $q$, and may be expressed as

\begin{equation}
e_{D}(q) = \frac{e_{raw}(q) - s_{m}(q)m_\text{MC}(q)f_{g}}{1 - f_{g}}~\text{,}
\label{eq:qg_eff}
\end{equation}
where $e_{raw}$ is the fraction of data events passing the QG requirement; $m_\text{MC}$ is the mistag rate for gluons, as measured in simulation; $s_{m}$ is a scale factor on the mistag rate in simulated jets to match the mistag rate measured in data; and $f_{g}$ is the fraction of gluon jets in the sample. We can write a similar expression for the mistag rate from

\begin{equation}
m_{D}(q) = \frac{m_{raw}(q) - s_{e}(q)e_\text{MC}(q)f_{q}}{1 - f_{q}}~\text{,}
\label{eq:qg_mis}
\end{equation}
where $m_{raw}$ is the fraction of data events meeting the QG requirements; $e_\text{MC}$ is the efficiency for quarks to pass the requirement, as measured in simulation; $s_{e}$ is a scale factor on the efficiency in simulated jets to match the mistag rate measured in data; and $f_{q}$ is the fraction of quark jets in the sample. Squared uncertainties on these quantities may be expressed as

\begin{align}
%\begin{equation}
  \sigma_{e}^{2}(q) &=
  \frac{1}{(1-f_{g})^{2}} \left[\frac{e_\text{raw}(1-e_\text{raw})}{N_{D}} 
    + (\sigma_{m}f_{g})^{2}\right] \notag \\
  &+ \sum_{X}\frac{\sigma_{X}^{2}}{\left[N_\text{MC}(1-f_{g})\right]^2} \times \notag \\
  &\left[(e+s_{m}m_\text{MC})(f_{g}-f_{g}^{X})\right. %\notag \\
   \left.+f_{q}^{X}(e_\text{MC}-e_{X})\right]^{2} ~\text{,}
\label{eq:efficiency-unc}
%\end{equation}
\end{align}
where $N_{D}$ and $N_\text{MC}$ are the number of data and simulated events, respectively, and where the $X$ represents the various simulated subsamples, and $\sigma_{m}$ is the uncertainty on the mistag rate, which may be expressed in an analogous fashion. The uncertainty includes a statistical uncertainty on the data, uncertainties on the mistag rate and efficiency, and uncertainties on the relative difference in the contributions from the simulation. We take the uncertainties on the normalizations of the $t\bar{t}$, single top-quark, diboson, $V$+jets, and $V+b$ jets to be 6\%, 10\%, 6\%, 10\%, and 40\%, respectively, based on the uncertainties in their production cross sections.

We measure the efficiency in the $t\bar{t}$ sample, where the fraction of gluon jets is small, and measure the mistag rate in the $W+2$ jets sample, where the gluon fraction is much larger and similar to the fraction in the $Z+2$ jets signal region. The efficiency, mistag rate, and their uncertainties are determined using an iterative procedure. We first calculate the mistag rate in data assuming that the efficiency in data equals the efficiency in simulation. We then calculate the efficiency in data assuming that value for the mistag rate and proceed to update the mistag rate assuming the new value for the efficiency from data. We observe rapid convergence on robust values for the efficiency and mistag rate.  Table~\ref{tab:qg_eff} shows the efficiency and mistag rate for the given requirement of minimum QG $>0.0$, measured in both data and MC simulation. The simulation underestimates the rate for quark jets to meet the jet QG requirement, while correctly predicting the observed mistag rate.

\begin{table*}[hbt]
\begin{center}
\begin{ruledtabular}
\begin{tabular}{lccc}
& MC & Data & MC revised jet QG requirement \\
& & & ($-1\sigma$,nom.,$+1\sigma$)\\ \hline
Efficiency &  0.241 & $0.295 \pm 0.034$ & $(-0.0325,-0.09,-0.14)$\\
Mistag rate &  0.088 & $0.087 \pm 0.027$ & $(0.09,-0.0175,-0.11)$\\
\end{tabular}
\end{ruledtabular}
\caption{Efficiency and mistag rates for the chosen jet QG requirements, as evaluated in data and MC simulation, along with the necessary change in the jet QG threshold for the simulation to model the proper rates and the uncertainties on them.}
\label{tab:qg_eff}
\end{center}
\end{table*}

We implement a correction to the MC simulation by varying the requirement on the minimum QG value in order to reproduce the efficiency and mistag rate observed in data. The uncertainties on these quantities are also obtained by varying the jet QG requirement. The alternate thresholds used for simulated quark and gluon jets are listed in Table~\ref{tab:qg_eff}.

\section{Signal extraction and results}
\label{sec:extraction}

We extract the number of signal events using a binned $\chi^{2}$-minimization fit to data, using the techniques described in Ref.~\cite{mclimit}. We create histogram templates for both signal and background samples. The templates, along with the uncertainties we assign to their normalization in the fit procedure, are listed below.
\begin{enumerate}
\item $ZV$ signal: We allow the normalization of the signal template to float unconstrained in the fit. We assume that each signal process contributes proportionally to its predicted SM cross section: 3.6 pb for \textit{ZW} and 1.5 pb for \textit{ZZ}~\cite{campbell}.

\item $Z$+jets: This is the largest background. We allow its normalization to float in the fit, unconstrained.

\item $Z$+$b$ jets: We constrain the normalization of this significant background to be within $\pm40\%$ of its nominal value.

\item $t\bar{t}$: We use a production cross section of $\sigma_{t\bar{t}}=7.5$~pb, and assign an uncertainty of $6.5\%$ to the normalization of this template, based on the theoretical cross-section uncertainty~\cite{ttbar-xsec}.

\item Misidentified leptons: We use the method described in Sec.~\ref{sec:selection} to construct templates for the contribution from jets mimicking one or two leptons. We assign an uncertainty of $50\%$ to the misidentification rate, based on studies using different trigger thresholds for the jet data used to obtain these rates.

\end{enumerate}

We perform a simultaneous fit to data using independent templates for each of three channels. For events passing the basic signal selection requirements described in Sec.~\ref{sec:selection}, we first construct a heavy-flavor tag (HF-tag) channel composed of events passing a minimum jet $b$ness requirement (jet $b$ness $> 0$), using the jet-$b$ness tagger~\cite{bness_nim}. For events failing this requirement, we then select events passing the minimum jet-QG value requirement described in Sec.~\ref{sec:QGvalue} to form a light-flavor tag (LF-tag) channel. Events failing these requirement are then placed in the third, untagged, channel, which has a lower signal fraction than the two tagged channels, but still includes a significant amount of signal due to the tight tagging requirements.

\begin{table*}[hbt]
\begin{center}
\begin{ruledtabular}
\begin{tabular}{llccccc}
Source & Channel & \textit{ZV} & $Z$+jets & $Z$+$b$ jets & $t\bar{t}$ & Mis-ID leptons \\ \hline
Cross section/norm. & All &  Unconstr. & Unconstr. & $\pm40\%$ & $\pm6.5\%$ & $\pm50\%$ \\
Jet energy res. & HF-tag & $\pm0.8\%$ & $\pm0.3\%$ & $\pm1.0\%$ & $\pm0.2\%$ &  \\
 & LF-tag & $\pm1.0\%$ & $\pm0.7\%$ & $\pm1.5\%$ & $\pm6.2\%$ &  \\
 & Untagged & $\pm0.6\%$ & $\pm0.9\%$ & $\pm0.7\%$ & $\pm1.1\%$ &  \\
Jet energy scale & HF-tag & $\pm4.0\%$ & $\pm4.4\%$ & $\pm3.8\%$ & $\pm4.0\%$ &  \\
 & LF-tag & $\pm1.5\%$ & $\pm0.3\%$ & $\pm0.6\%$ & $\pm 3.0\%$ &  \\
 & Untagged & $\pm1.9\%$ & $\pm5.7\%$ & $\pm3.8\%$ & $\pm1.9\%$ &  \\
$Q^{2}$ & All & none & Shape only & Shape only & none &  \\
ISR/FSR & All & Shape only  & none & none & none &  \\
$b$ness tag & HF-tag &  $\pm 7.8\%$ & $\pm7.8\%$ & $\pm9.2\%$ & $\pm7.6\%$ &  \\
 & LF-tag &  $\pm0.2\%$ & $\pm0.0\%$ & $\pm1.2\%$ & $\pm2.8\%$ &  \\
 & Untagged &  $\pm0.4\%$ & $\pm0.1\%$ & $\pm1.8\%$ & $\pm4.5\%$ &  \\
QG tag & LF-tag & $\pm10\%$  & $\pm16\%$ & $\pm2.0\%$ & $\pm15\%$ &  \\
 & Untagged & $\pm4.3\%$  & $\pm3.5\%$ & $\pm2.0\%$ & $\pm2.0\%$ &  \\
Lepton energy scale & All & $\pm0.5\%$  & $\pm0.5\%$ & $\pm0.5\%$ & $\pm 1.5\%$ &  \\
Lepton energy res. & All & $\pm0.1\%$  & $\pm0.1\%$ & $\pm0.0\%$ & $\pm2.7\%$ &  \\ 
\end{tabular}
\end{ruledtabular}
\caption{Summary of the systematic uncertainties considered in the fit of the dijet mass distribution. Uncertainties that change both the shape and rate of templates used in the fit are treated in a correlated fashion.}
\label{tab:systematics}
\end{center}
\end{table*}

Additional systematic uncertainties on both the normalization and shapes of the templates used in the fit are also considered. We estimate uncertainties due to mismodeling between data and MC simulation in the jet energy scale (as described in Sec.~\ref{sec:jesQG_systematics}) and the jet energy resolution, the modeling of the tagging variables, and the lepton energy scale and resolution. Additional shape uncertainties on the $Z$+jets backgrounds are considered by increasing and decreasing the renormalization and factorization scale, $Q^{2}$, from the default value in the simulation of $m^{2}_{Z} + p^{2}_{T,Z}$. We also consider the effect on the shape of the dijet invariant mass when increasing or decreasing initial- and final-state QCD radiation (ISR/FSR) in the $ZV$ signal model. These systematic uncertainties, along with the normalization constraints described above, are treated as nuisance parameters in the fit, and are included in the $\chi^{2}$-minimization procedure \cite{mclimit}. They are summarized in Table~\ref{tab:systematics}.

\begin{figure*}[htbp]
\centering
{\includegraphics[width=0.9\textwidth]{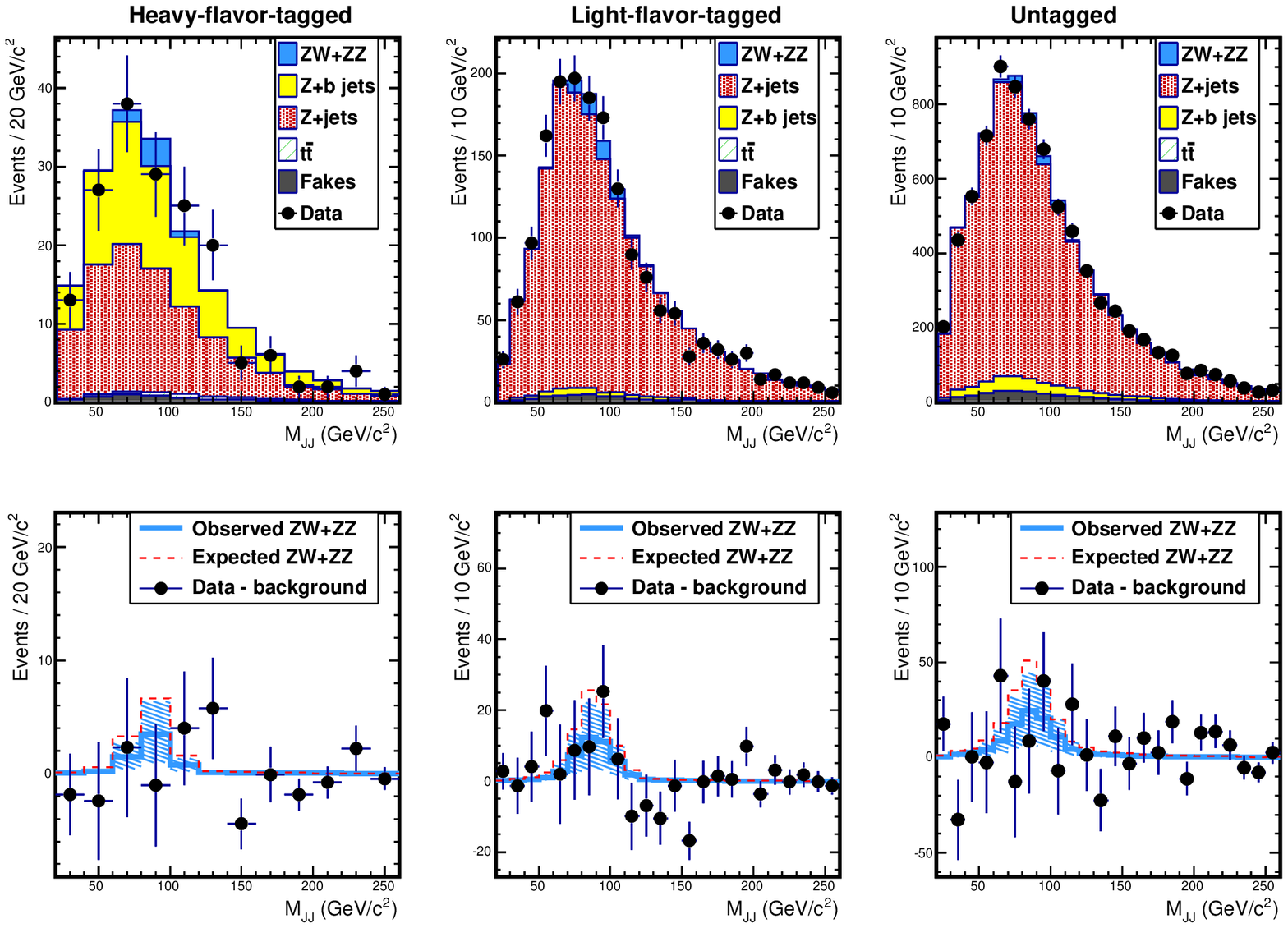}}
\caption{Invariant dijet mass distributions with fit results overlaid for the $ZW+ZZ$ process in the dilepton+dijet selection in the heavy-flavor--tagged channel (left panels), light-flavor-tagged channel (center panels), and untagged channel (right panels). The top row shows the output from the fit compared to the data, while the bottom row shows the background subtracted from data, compared to the expected (dashed line) and fitted (solid line, with uncertainties in bands) signal contributions.}
\label{fig:results}
\end{figure*}

\begin{table*}[hbt]
\begin{center}
\begin{ruledtabular}
\begin{tabular}{lccc}
Process & $N_{events}$, HF-tag & $N_{events}$, LF-tag & $N_{events}$, Untagged \\ \hline
$Z$+jets & $91.9 \pm 8.3$ & $1\,605 \pm 50$ & $7\,200 \pm 600$ \\
$Z$+$b$ jets & $71 \pm 14$ & $37 \pm 10$ & $360 \pm 100$ \\
%$t\bar{t}$ & $3.18 \pm 0.35$ & $0.71 \pm 0.07$& $5.26 \pm 0.42$ \\
$t\bar{t}$ & $3.2 \pm 0.4$ & $0.7 \pm 0.1$& $5.3 \pm 0.4$ \\
Misidentified leptons & $4.6 \pm 2.3$ & $39 \pm 20$ & $270 \pm 140$ \\
\hline
Total background & $171 \pm 14$ & $1\,681 \pm 36$ & $7\,840 \pm 600$ \\
$ZW+ZZ$ & $6.3 \pm 4.4$ & $45 \pm 30$ & $106 \pm 72$ \\
\hline
Total events & $177 \pm 14$ & $1\,726 \pm 40$ & $7\,940 \pm 610$ \\
Data events & 172 & $1\,724$ & $7\,950$ \\
%\bf{KS Probability} & & & \\
%Final Fit & 0.97 & 0.61 & 0.99 \\
%SM $\sigma_{WZ+ZZ}$ & 0.96 & 0.68 & 0.82\\
%No Signal & 0.95 & 0.27 & 0.92\\
\end{tabular}
\end{ruledtabular}
\caption{Number of events in each class from the best fit to the data.}
\label{tab:results}
\end{center}
\end{table*}

Figure~\ref{fig:results} shows the dijet mass distributions in data with the fit results overlaid. Table~\ref{tab:results} shows the number of events of each class determined by the fit. We fit for approximately $50\%$ of the expected signal normalization and observe good agreement between data and simulation in the final fit for each of the three channels, with a total $\chi^{2}/$d.o.f$ = 59.8/55$.

\begin{figure*}[htb]
	\centering
	\includegraphics[width=0.7\textwidth]{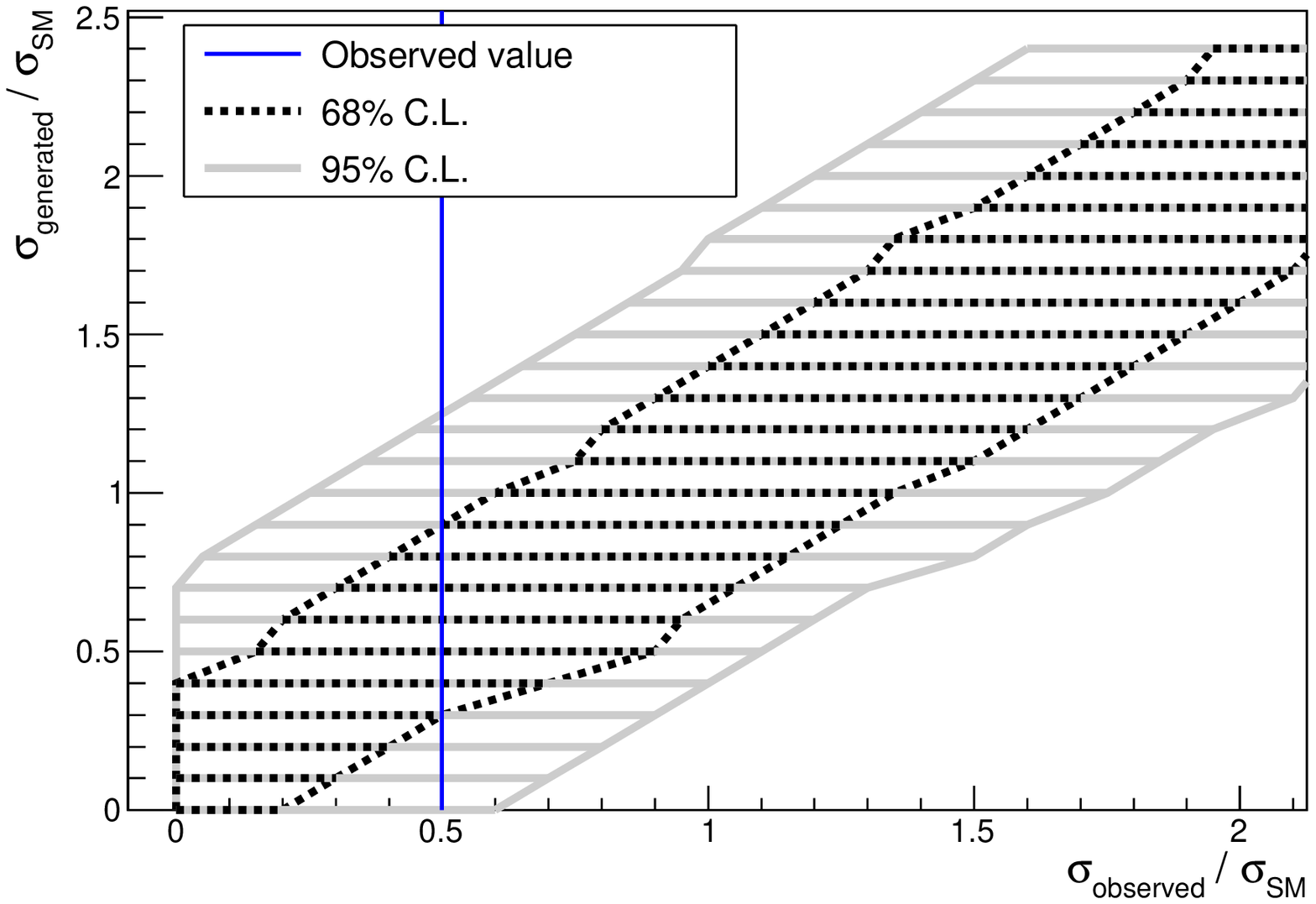}
\caption{Confidence bands showing the expected range of observed cross sections as a function of the true cross section, with 68\% C.L. (black dashed region) and 95\% C.L. (solid gray region).}
\label{fig:FC}
\end{figure*}

We do not see significant evidence for $ZW+ZZ$ production in this decay channel. Hence, we set upper limits on the production cross section using likelihood-ratio ordering~\cite{FC}, where we analyze the distribution of observed cross sections in pseudoexperiments generated with a variety of scale factors on the input signal cross section. When generating pseudoexperiments, we consider additional systematic uncertainties that affect the acceptance, assigning a $2\%$ uncertainty on the signal template from limited knowledge of the from parton distribution functions, and $2.5\%$ and $6\%$ uncertainties due to the uncertainties on the lepton-scale-factor and integrated luminosity, respectively. The set of input cross sections in the pseudoexperiments range from 0.0 to 2.9 times the expected cross section, with a step size of $0.1$.

Figure~\ref{fig:FC} shows the resulting confidence band. Using the $1\sigma$ bands, we determine $\sigma (\ppbar \rightarrow ZW+ZZ) = 2.5^{+2.0}_{-1.0}$~pb, compared to the standard model prediction of $\sigma_\text{SM} = 5.1$~pb. We do not exclude the no-signal hypothesis, and establish a limit of $\sigma_{ZW+ZZ} < 6.1$~pb ($1.25 \times \sigma_\text{SM}$) at the 95\% C.L.

\section{Summary}
In this paper we describe a search for \textit{ZW} and \textit{ZZ} boson pair-production from a final state with two charged, high-transverse-momentum electrons or muons and two hadronic jets. We increase the sensitivity by tagging events with jets likely originating from heavy- and light-flavor quarks and classifying them in separate analysis channels using neural-network-based taggers. These taggers benefit from the large sample of events containing top quarks collected by CDF, allowing a data-driven estimate of the efficiency and mistag rates for jets passing tagging requirements. We also improve the modeling of the Monte Carlo simulations, especially those that describe the $Z +$~jets background, by deriving and incorporating improved energy corrections for simulated jets to more accurately reproduce the phenomenology of jets originating from quarks and gluons in the data.

Using the full CDF Run {\rm II} proton-antiproton collisions data set, which corresponds to an integrated luminosity of $8.9$~fb$^{-1}$, we fit for the normalization of $ZW, ZZ \rightarrow \ell^{+}\ell^{-}+q\bar{q'}$ events using the dijet invariant mass distribution. We incorporate many of the systematic uncertainties associated with the modeling of signal and background processes as nuisance parameters in the dijet mass fit. We measure a cross section of $\sigma_{ZW+ZZ} = 2.5^{+2.0}_{-1.0}~\text{pb,}$ which is nonzero at the $1.75\sigma$ level of significance. We also obtain a limit on the cross section of $\sigma_{ZW+ZZ} < 6.1~\text{pb at the 95\% C.~L.}$

\begin{acknowledgments}
We thank the Fermilab staff and the technical staffs of the
participating institutions for their vital contributions. This work
was supported by the U.S. Department of Energy and National Science
Foundation; the Italian Istituto Nazionale di Fisica Nucleare; the
Ministry of Education, Culture, Sports, Science and Technology of
Japan; the Natural Sciences and Engineering Research Council of
Canada; the National Science Council of the Republic of China; the
Swiss National Science Foundation; the A.P. Sloan Foundation; the
Bundesministerium f\"ur Bildung und Forschung, Germany; the Korean
World Class University Program, the National Research Foundation of
Korea; the Science and Technology Facilities Council and the Royal
Society, United Kingdom; the Russian Foundation for Basic Research;
the Ministerio de Ciencia e Innovaci\'{o}n, and Programa
Consolider-Ingenio 2010, Spain; the Slovak R\&D Agency; the Academy
of Finland; the Australian Research Council (ARC); and the EU community
Marie Curie Fellowship Contract No. 302103.

\end{acknowledgments}

\newpage
\bibliography{prd}

\begin{thebibliography}{26}
\expandafter\ifx\csname natexlab\endcsname\relax\def\natexlab#1{#1}\fi
\expandafter\ifx\csname bibnamefont\endcsname\relax
  \def\bibnamefont#1{#1}\fi
\expandafter\ifx\csname bibfnamefont\endcsname\relax
  \def\bibfnamefont#1{#1}\fi
\expandafter\ifx\csname citenamefont\endcsname\relax
  \def\citenamefont#1{#1}\fi
\expandafter\ifx\csname url\endcsname\relax
  \def\url#1{\texttt{#1}}\fi
\expandafter\ifx\csname urlprefix\endcsname\relax\def\urlprefix{URL }\fi
\providecommand{\bibinfo}[2]{#2}
\providecommand{\eprint}[2][]{\url{#2}}

\bibitem[{\citenamefont{Campbell and Ellis}(1999)}]{campbell}
\bibinfo{author}{\bibfnamefont{J.~M.} \bibnamefont{Campbell}} \bibnamefont{and}
  \bibinfo{author}{\bibfnamefont{R.~K.} \bibnamefont{Ellis}},
  \bibinfo{journal}{Phys. Rev. D} \textbf{\bibinfo{volume}{60}},
  \bibinfo{pages}{113006} (\bibinfo{year}{1999}).

\bibitem[{\citenamefont{Hagiwara et~al.}(1987)\citenamefont{Hagiwara, Peccei,
  Zeppenfeld, and Hikasa}}]{hagiwara}
\bibinfo{author}{\bibfnamefont{K.}~\bibnamefont{Hagiwara}},
  \bibinfo{author}{\bibfnamefont{R.}~\bibnamefont{Peccei}},
  \bibinfo{author}{\bibfnamefont{D.}~\bibnamefont{Zeppenfeld}},
  \bibnamefont{and} \bibinfo{author}{\bibfnamefont{K.}~\bibnamefont{Hikasa}},
  \bibinfo{journal}{Nucl. Phys.} \textbf{\bibinfo{volume}{B282}},
  \bibinfo{pages}{253} (\bibinfo{year}{1987}).

\bibitem[{\citenamefont{Kober et~al.}(2007)\citenamefont{Kober, Koch, and
  Bleicher}}]{kober}
\bibinfo{author}{\bibfnamefont{M.}~\bibnamefont{Kober}},
  \bibinfo{author}{\bibfnamefont{B.}~\bibnamefont{Koch}}, \bibnamefont{and}
  \bibinfo{author}{\bibfnamefont{M.}~\bibnamefont{Bleicher}},
  \bibinfo{journal}{Phys. Rev. D} \textbf{\bibinfo{volume}{76}},
  \bibinfo{pages}{125001} (\bibinfo{year}{2007}).

\bibitem[{\citenamefont{Eichten et~al.}(2011)\citenamefont{Eichten, Lane, and
  Martin}}]{technicolor}
\bibinfo{author}{\bibfnamefont{E.~J.} \bibnamefont{Eichten}},
  \bibinfo{author}{\bibfnamefont{K.}~\bibnamefont{Lane}}, \bibnamefont{and}
  \bibinfo{author}{\bibfnamefont{A.}~\bibnamefont{Martin}},
  \bibinfo{journal}{Phys. Rev. Lett.} \textbf{\bibinfo{volume}{106}},
  \bibinfo{pages}{251803} (\bibinfo{year}{2011}).

\bibitem[{\citenamefont{Aaltonen et~al.}(2009{\natexlab{a}})}]{metjj_prl}
\bibinfo{author}{\bibfnamefont{T.}~\bibnamefont{Aaltonen}} \bibnamefont{et~al.}
  (\bibinfo{collaboration}{CDF Collaboration}), \bibinfo{journal}{Phys. Rev.
  Lett.} \textbf{\bibinfo{volume}{103}}, \bibinfo{pages}{091803}
  (\bibinfo{year}{2009}{\natexlab{a}}).

\bibitem[{\citenamefont{Aaltonen et~al.}(2010{\natexlab{a}})}]{lmetjj_prl_ME}
\bibinfo{author}{\bibfnamefont{T.}~\bibnamefont{Aaltonen}} \bibnamefont{et~al.}
  (\bibinfo{collaboration}{CDF Collaboration}), \bibinfo{journal}{Phys. Rev.
  Lett.} \textbf{\bibinfo{volume}{104}}, \bibinfo{pages}{101801}
  (\bibinfo{year}{2010}{\natexlab{a}}).

\bibitem[{\citenamefont{Aaltonen et~al.}(2011)}]{lmetjj_prl_Mjj}
\bibinfo{author}{\bibfnamefont{T.}~\bibnamefont{Aaltonen}} \bibnamefont{et~al.}
  (\bibinfo{collaboration}{CDF Collaboration}), \bibinfo{journal}{Phys. Rev.
  Lett.} \textbf{\bibinfo{volume}{106}}, \bibinfo{pages}{171801}
  (\bibinfo{year}{2011}).

\bibitem[{\citenamefont{Abazov et~al.}(2012)}]{D0_lmetjj}
\bibinfo{author}{\bibfnamefont{V.~M.} \bibnamefont{Abazov}}
  \bibnamefont{et~al.} (\bibinfo{collaboration}{D0 Collaboration}),
  \bibinfo{journal}{Phys. Rev. Lett.} \textbf{\bibinfo{volume}{108}},
  \bibinfo{pages}{181803} (\bibinfo{year}{2012}).

\bibitem[{\citenamefont{Aaltonen et~al.}(2012)}]{metbb_prd}
\bibinfo{author}{\bibfnamefont{T.}~\bibnamefont{Aaltonen}} \bibnamefont{et~al.}
  (\bibinfo{collaboration}{CDF Collaboration}), \bibinfo{journal}{Phys. Rev. D}
  \textbf{\bibinfo{volume}{85}}, \bibinfo{pages}{012002}
  (\bibinfo{year}{2012}).

\bibitem[{\citenamefont{Ketchum}(2012)}]{Ketchum_thesis_prl}
\bibinfo{author}{\bibfnamefont{W.}~\bibnamefont{Ketchum}},
  \bibinfo{journal}{Ph.D. thesis, University of Chicago.
  FERMILAB-THESIS-2012-36}  (\bibinfo{year}{2012}).

\bibitem[{coo()}]{coordinateSystem}
\bibinfo{note}{The coordinate system used to describe the detector has its
  origin located at the detector's center. The proton beam direction is defined
  as the $+z$ direction. The polar angle, $\theta$, is measured from the origin
  with respect to the $z$ axis, and $\phi$ is the azimuthal angle.
  Pseudorapidity, transverse energy, and transverse momentum are defined as
  $\eta$=$-\ln\tan(\theta/2)$, $E_{T}$=$E\sin\theta$, and
  $p_{T}$=$p\sin\theta$, respectively. The rectangular coordinates $x$ and $y$
  point radially outward and vertically upward from the Tevatron ring,
  respectively.}

\bibitem[{\citenamefont{Abulencia et~al.}(2007)}]{CDF_detect_A}
\bibinfo{author}{\bibfnamefont{A.}~\bibnamefont{Abulencia}}
  \bibnamefont{et~al.} (\bibinfo{collaboration}{CDF Collaboration}),
  \bibinfo{journal}{J. Phys. G} \textbf{\bibinfo{volume}{34}},
  \bibinfo{pages}{2457} (\bibinfo{year}{2007}).

\bibitem[{\citenamefont{Abe et~al.}(1992)}]{jetclu}
\bibinfo{author}{\bibfnamefont{F.}~\bibnamefont{Abe}} \bibnamefont{et~al.}
  (\bibinfo{collaboration}{CDF Collaboration}), \bibinfo{journal}{Phys. Rev. D}
  \textbf{\bibinfo{volume}{45}}, \bibinfo{pages}{1448} (\bibinfo{year}{1992}).

\bibitem[{met()}]{metdef}
\bibinfo{note}{We define the missing transverse momentum $\vec{\mett}$$\equiv
  -\sum_i E_{\rm T}^i {\bf n}_i$, where ${\bf n}_i$ is the unit vector in the
  azimuthal plane that points from the beamline to the $i$th calorimeter tower.
  We call the magnitude of this vector the missing transverse energy, \mett.}

\bibitem[{\citenamefont{Mangano et~al.}(2003)\citenamefont{Mangano, Moretti,
  Piccinini, Pittau, and Polosa}}]{alpgen}
\bibinfo{author}{\bibfnamefont{M.~L.} \bibnamefont{Mangano}},
  \bibinfo{author}{\bibfnamefont{M.}~\bibnamefont{Moretti}},
  \bibinfo{author}{\bibfnamefont{F.}~\bibnamefont{Piccinini}},
  \bibinfo{author}{\bibfnamefont{R.}~\bibnamefont{Pittau}}, \bibnamefont{and}
  \bibinfo{author}{\bibfnamefont{A.~D.} \bibnamefont{Polosa}},
  \bibinfo{journal}{J. High Energy Phys.}
  \textbf{\bibinfo{volume}{\textnormal{07}}}, \bibinfo{pages}{001}
  (\bibinfo{year}{2003}).

\bibitem[{\citenamefont{Sjostrand et~al.}(2006)\citenamefont{Sjostrand, Mrenna,
  and Skands}}]{pythia}
\bibinfo{author}{\bibfnamefont{T.}~\bibnamefont{Sjostrand}},
  \bibinfo{author}{\bibfnamefont{S.}~\bibnamefont{Mrenna}}, \bibnamefont{and}
  \bibinfo{author}{\bibfnamefont{P.~Z.} \bibnamefont{Skands}},
  \bibinfo{journal}{J. High Energy Phys.}
  \textbf{\bibinfo{volume}{\textnormal{05}}}, \bibinfo{pages}{026}
  (\bibinfo{year}{2006}).

\bibitem[{\citenamefont{Aaltonen et~al.}(2009{\natexlab{b}})}]{ZbPRD}
\bibinfo{author}{\bibfnamefont{T.}~\bibnamefont{Aaltonen}} \bibnamefont{et~al.}
  (\bibinfo{collaboration}{CDF Collaboration}), \bibinfo{journal}{Phys. Rev. D}
  \textbf{\bibinfo{volume}{79}}, \bibinfo{pages}{052008}
  (\bibinfo{year}{2009}{\natexlab{b}}).

\bibitem[{\citenamefont{Bhatti et~al.}(2006)}]{cdf_JES}
\bibinfo{author}{\bibfnamefont{A.}~\bibnamefont{Bhatti}} \bibnamefont{et~al.},
  \bibinfo{journal}{Nucl. Instrum. Methods A} \textbf{\bibinfo{volume}{566}},
  \bibinfo{pages}{375 } (\bibinfo{year}{2006}).

\bibitem[{\citenamefont{Paramonov et~al.}(2010)\citenamefont{Paramonov,
  Canelli, D'Onofrio, Frisch, and Mrenna}}]{cdf_JES_precision_limits}
\bibinfo{author}{\bibfnamefont{A.}~\bibnamefont{Paramonov}},
  \bibinfo{author}{\bibfnamefont{F.}~\bibnamefont{Canelli}},
  \bibinfo{author}{\bibfnamefont{M.}~\bibnamefont{D'Onofrio}},
  \bibinfo{author}{\bibfnamefont{H.}~\bibnamefont{Frisch}}, \bibnamefont{and}
  \bibinfo{author}{\bibfnamefont{S.}~\bibnamefont{Mrenna}},
  \bibinfo{journal}{Nucl.~Instrum.~Methods A} \textbf{\bibinfo{volume}{622}},
  \bibinfo{pages}{698} (\bibinfo{year}{2010}).

\bibitem[{\citenamefont{Hoecker et~al.}(2007)\citenamefont{Hoecker, Speckmayer,
  Stelzer, Therhaag, von Toerne, and Voss}}]{tmva}
\bibinfo{author}{\bibfnamefont{A.}~\bibnamefont{Hoecker}},
  \bibinfo{author}{\bibfnamefont{P.}~\bibnamefont{Speckmayer}},
  \bibinfo{author}{\bibfnamefont{J.}~\bibnamefont{Stelzer}},
  \bibinfo{author}{\bibfnamefont{J.}~\bibnamefont{Therhaag}},
  \bibinfo{author}{\bibfnamefont{E.}~\bibnamefont{von Toerne}},
  \bibnamefont{and} \bibinfo{author}{\bibfnamefont{H.}~\bibnamefont{Voss}},
  \bibinfo{journal}{PoS} \textbf{\bibinfo{volume}{ACAT}}, \bibinfo{pages}{040}
  (\bibinfo{year}{2007}).

\bibitem[{mTd()}]{mTdef}
\bibinfo{note}{We define the transverse mass as $m_{T} =
  \sqrt{2p_{T}^\text{lepton}\mett\left(1 -
  \cos{\Delta\phi_{\ell,\mett}}\right)}$, where $\Delta\phi_{\ell,\mett}$ is
  the difference in $\phi$ in the direction of the lepton $\vec{p_{T}}$ and
  \mettvec vectors.}

\bibitem[{\citenamefont{Alwall et~al.}(2007)\citenamefont{Alwall, Demin,
  de~Visscher, Frederix, Herquet, Maltoni, Plehn, Rainwater, and
  Stelzer}}]{madgraph}
\bibinfo{author}{\bibfnamefont{J.}~\bibnamefont{Alwall}},
  \bibinfo{author}{\bibfnamefont{P.}~\bibnamefont{Demin}},
  \bibinfo{author}{\bibfnamefont{S.}~\bibnamefont{de~Visscher}},
  \bibinfo{author}{\bibfnamefont{R.}~\bibnamefont{Frederix}},
  \bibinfo{author}{\bibfnamefont{M.}~\bibnamefont{Herquet}},
  \bibinfo{author}{\bibfnamefont{F.}~\bibnamefont{Maltoni}},
  \bibinfo{author}{\bibfnamefont{T.}~\bibnamefont{Plehn}},
  \bibinfo{author}{\bibfnamefont{D.~L.} \bibnamefont{Rainwater}},
  \bibnamefont{and} \bibinfo{author}{\bibfnamefont{T.}~\bibnamefont{Stelzer}},
  \bibinfo{journal}{J. High Energy Phys.}
  \textbf{\bibinfo{volume}{\textnormal{09}}}, \bibinfo{pages}{028}
  (\bibinfo{year}{2007}).

\bibitem[{\citenamefont{Freeman et~al.}(2012)\citenamefont{Freeman, Lewis,
  Ketchum, Poprocki, Pronko, Rusu, and Wittich}}]{bness_nim}
\bibinfo{author}{\bibfnamefont{J.}~\bibnamefont{Freeman}},
  \bibinfo{author}{\bibfnamefont{J.}~\bibnamefont{Lewis}},
  \bibinfo{author}{\bibfnamefont{W.}~\bibnamefont{Ketchum}},
  \bibinfo{author}{\bibfnamefont{S.}~\bibnamefont{Poprocki}},
  \bibinfo{author}{\bibfnamefont{A.}~\bibnamefont{Pronko}},
  \bibinfo{author}{\bibfnamefont{V.}~\bibnamefont{Rusu}}, \bibnamefont{and}
  \bibinfo{author}{\bibfnamefont{P.}~\bibnamefont{Wittich}},
  \bibinfo{journal}{Nucl. Instrum. Methods A} \textbf{\bibinfo{volume}{663}},
  \bibinfo{pages}{37} (\bibinfo{year}{2012}).

\bibitem[{\citenamefont{Aaltonen et~al.}(2010{\natexlab{b}})}]{mclimit}
\bibinfo{author}{\bibfnamefont{T.}~\bibnamefont{Aaltonen}} \bibnamefont{et~al.}
  (\bibinfo{collaboration}{CDF Collaboration}), \bibinfo{journal}{Phys. Rev. D}
  \textbf{\bibinfo{volume}{82}}, \bibinfo{pages}{112005}
  (\bibinfo{year}{2010}{\natexlab{b}}).

\bibitem[{\citenamefont{Langenfeld et~al.}(2009)\citenamefont{Langenfeld, Moch,
  and Uwer}}]{ttbar-xsec}
\bibinfo{author}{\bibfnamefont{U.}~\bibnamefont{Langenfeld}},
  \bibinfo{author}{\bibfnamefont{S.}~\bibnamefont{Moch}}, \bibnamefont{and}
  \bibinfo{author}{\bibfnamefont{P.}~\bibnamefont{Uwer}},
  \bibinfo{journal}{Phys. Rev. D} \textbf{\bibinfo{volume}{80}},
  \bibinfo{pages}{054009} (\bibinfo{year}{2009}).

\bibitem[{\citenamefont{Feldman and Cousins}(1998)}]{FC}
\bibinfo{author}{\bibfnamefont{G.~J.} \bibnamefont{Feldman}} \bibnamefont{and}
  \bibinfo{author}{\bibfnamefont{R.~D.} \bibnamefont{Cousins}},
  \bibinfo{journal}{Phys. Rev. D} \textbf{\bibinfo{volume}{57}},
  \bibinfo{pages}{3873} (\bibinfo{year}{1998}).

\end{thebibliography}

\end{document}